# A RELUCTANT ADDITIVE MODEL FRAMEWORK FOR INTERPRETABLE NONLINEAR INDIVIDUALIZED TREATMENT RULES

BY JACOB M. MARONGE[1], JARED D. HULING[2,*] AND GUANHUA CHEN[3,†]

[1]*Department of Biostatistics, University of Texas MD Anderson Cancer Center.*

[2]*Division of Biostatistics, University of Minnesota.* *huling@umn.edu*

[3]*Department of Biostatistics and Medical Informatics, University of Wisconsin-Madison.* †gchen25@wisc.edu*

Individualized treatment rules (ITRs) for treatment recommendation is an important topic for precision medicine as not all beneficial treatments work well for all individuals. Interpretability is a desirable property of ITRs, as it helps practitioners make sense of treatment decisions, yet there is a need for ITRs to be flexible to effectively model complex biomedical data for treatment decision making. Many ITR approaches either focus on linear ITRs, which may perform poorly when true optimal ITRs are nonlinear, or black-box nonlinear ITRs, which may be hard to interpret and can be overly complex. This dilemma indicates a tension between interpretability and accuracy of treatment decisions. Here we propose an additive model-based nonlinear ITR learning method that balances interpretability and flexibility of the ITR. Our approach aims to strike this balance by allowing both linear and nonlinear terms of the covariates in the final ITR. Our approach is parsimonious in that the nonlinear term is included in the final ITR only when it substantially improves the ITR performance. To prevent overfitting, we combine cross-fitting and a specialized information criterion for model selection. Through extensive simulations, we show that our methods are data-adaptive to the degree of nonlinearity and can favorably balance ITR interpretability and flexibility. We further demonstrate the robust performance of our methods with an application to a cancer drug sensitive study.

**1. Introduction.** Machine learning tools have been developed extensively for health care applications and have the opportunity to deeply impact human lives (Rudin, 2019). A major area where machine learning has shown promise is in the development of individualized treatment rules (ITRs), which are used to help personalize and tailor health care decisions based on the individual characteristics of patients. Many of these approaches are black boxes in that the model-fitting procedure does not yield results that are readily interpretable to subject-matter experts or patients. The lack of transparency and accountability of black-box models may have severe consequences, such as the misdiagnosis of diseases (Rudin, 2019) or incorrect interpretation of the modeled relationships between factors. This is particularly evident when identifying ITRs. As ITRs (and the methods used to identify them) are directly related to medical treatment decisions (Vayena et al., 2018), there is a need for ITR learning methods that not only yield accurate recommendation, but also achieve a level of interpretability. Moreover, intepretable ITRs could be more accessible for stakeholders, including physicians and patients, and easier to implement in medical practice.

The goal of this paper is to develop an approach for estimating ITRs that are both accurate and interpretable. There is often a trade-off between model performance (predictive accuracy) and model complexity (interpretability) (Hastie et al., 2009; Murdoch et al., 2019) and we aim to strike a balance along this trade-off. Simpler models tend to be easier to interpret, but may fail to fully extract signals from rich, heterogeneous data, as is common in studies of human health. On the other hand, complex black-box models may perform well, but can be challenging to interpret and can have issues with over-fitting. Most existing methods either result in linear ITRs (Qian and Murphy, 2011; Wallace and Moodie, 2015; Chen et al., 2017), which may perform poorly when the underlying true ITR is nonlinear, or black-box ITRs (e.g., the ones based on Random Forests (Zhu et al., 2017; Wager and Athey, 2018), the ones based on deep neural networks (Liang et al., 2018; Mi et al., 2019), or the ones based on supporting vector machines with nonlinear kernel (Zhao et al., 2012; Zhou et al., 2017)), which may be difficult to interpret. In order to achieve a balance between these two problems, in a similar spirit to the reluctant additive model (RAM) of Tay and Tibshirani (2020), we propose a reluctant additive model for identifying ITRs. The key principle of the reluctant additive model is to allow for nonlinearity, but to shrink preferentially towards a linear term unless warranted by the data. We refer to our approach as Reluctant Additive Individualized Treatment Rules (RAITR) as it adheres to the reluctant additive principle.

Classical additive models require the user to choose *a priori* which covariates require a nonlinear functional form and which should be linear, which is a limitation especially in high-dimensional scenarios. As a result, classical additive models can be overly complex. Several approaches in the literature have explicitly focused on handling high dimensionality for additive models by inducing sparsity with penalties, such as the approaches of Lin and Zhang (2006), Ravikumar et al. (2009), and Meier et al. (2009). However, these approaches either shrink the entire functional form of the effect of a covariate to zero or otherwise include a fully nonlinear term. The fused LASSO additive model approach of Petersen et al. (2016) allows for shrinkage towards piece-wise constant effects, enabling more parsimonious selected models than approaches that only focus on selection, which either include or exclude a nonlinear term in its entirety. The sparse partially linear additive model of Petersen and Witten (2019) is able to induce sparsity for GAMs and can shrink towards linear functions, however, we adapted and extend RAM for learning ITRs instead due to its simple computational nature, which will be explained below.

In the context of ITRs, the work of Park et al. (2022) has extended the notion of sparse additive models to handle sparse additive ITRs, which inherits the limitations of the above sparse additive methods. In contrast, RAMs decompose the effect of a covariate on the outcome into a linear and nonlinear component and prioritize the linear terms in estimation, encouraging more aggressive shrinkage towards a parsimonious model while also allowing selection of individual variables into the model. This is achieved by fitting the linear terms first, as well as penalizing the nonlinear terms based on their complexity and only allowing their inclusion when they substantially improve the model performance above and beyond a linear form. As ITR estimation involves treatment-covariate interactions, using a highly flexible model to estimate the interactions could lead to overfitting. As such, the ability of our framework to default to linear contributions to the ITR is critical and beneficial in practice. Moreover, the variable selection ability of our framework makes it more compelling for learning ITRs with high dimensional covariates compared to approaches that do not incorporate explicit variable selection such as causal (random) forests (Wager and Athey, 2018) or weighted support vector machines (e.g. with a Gaussian kernel) (Zhou et al., 2017). We believe that our RAITR method has promise for effective use in practice as it offers the possibility for handling complex forms of heterogeneity of effects, while encouraging parsimony and interpretable effects through shrinkage towards linear ITRs.

The reluctant additive principle, while advantageous in many ways, also creates unique challenges in the setting of ITR estimation. Since the reluctant additive principle is inherently multi-staged, care is required to prevent over-fitting. We combine cross-fitting, to help prevent double-dipping outcome information, with use of the concordance information criteria (CIC; Shi et al. (2021)) to perform the final ITR model selection. As we demonstrate empirically, implementing the CIC performs quite well at identifying complex treatment rules while simultaneously preventing over-fitting.

This paper is structured as follows. In Section 2, we define notation and assumptions to estimate ITRs in the context of observational data, as well as describe the model-fitting procedure for RAITR. In Section 3, we conduct a series of simulations to compare ITRs estimated from our approach to simple linear ITRs and complicated black-box nonlinear ITRs. In Section 4, we apply RAITR to learn ITRs for cancer treatment using the Genomics of Drug Sensitivity in Cancer (GDSC) study (Iorio et al., 2016), where the data are high-dimensional and of low to moderate signal. We demonstrate that for studies like GDSC it is useful for a method to be flexible enough to detect nonlinear ITRs, but also pick simpler models when required. Finally, we conclude with a discussion in Section 5.

## 2. Methods.

2.1. *Notation and assumptions.* We refer to our treatment of interest as an exposure to emphasize that we are working in an observational context. Each independent observation is composed of a triplet $(Y, A, \mathbf{X})$, where $Y$ is the observed response, $A$ is a binary exposure taking values $1$ or $-1$, and $\mathbf{X}$ is a $p$-dimensional vector of pre-exposure covariates of interest. For ease of notation, we make the assumption that a larger value of $Y$ corresponds to a more desirable outcome. Adopting the potential outcomes framework of Rubin (2005), we denote $Y(a)$ as the potential outcome if the exposure were set to level $a$. We then define

$$(2.1) \qquad \Delta(\mathbf{x}) \equiv E[Y(1)|\mathbf{X} = \mathbf{x}] - E[Y(-1)|\mathbf{X} = \mathbf{x}]$$

as the conditional (causal) average treatment effect (CATE) given $\mathbf{x}$.

In many real-world scenarios, the CATE varies as a function of covariates, which is a form of a "heterogeneous treatment effect" (HTE). The CATE is closely related to individualized treatment rules (ITR) in the sense that

$$d^*(\mathbf{x}) = \text{sign}\left\{\Delta(\mathbf{x})\right\}$$

is an optimal ITR. As such, an optimal ITR can be estimated by modeling $\Delta(\cdot)$. Our modeling strategy focuses on estimating a flexible ITR through a flexible parameterization of the CATE, while maintaining the qualitative and visual interpretability of the resulting ITR.

To relate $\Delta(\mathbf{x})$ to observable quantities, we impose the following standard assumptions for causal identification:

(i) Consistency: $Y = \mathbb{1}(A = 1)Y(1) + \mathbb{1}(A = -1)Y(-1)$
(ii) Strong ignorability: $A \perp\!\!\!\perp (Y(1), Y(-1))|\mathbf{X}$
(iii) Positivity: $0 < \pi(\mathbf{x}, a) < 1$ for all $\mathbf{x} \in \mathcal{X}$ for $a \in \{-1, 1\}$, where $\pi(\mathbf{x}, a) = \Pr(A = a|\mathbf{X} = \mathbf{x})$.

With these assumptions, (2.1) is equivalent to

$$(2.2) \qquad \Delta(\mathbf{x}) = E(Y|A = 1, \mathbf{X} = \mathbf{x}) - E(Y|A = -1, \mathbf{X} = \mathbf{x}) \ .$$

We can highlight the role of the conditional treatment effect by re-expressing the conditional mean function as

$$(2.3) \qquad E(Y|A = a, \mathbf{X} = \mathbf{x}) = \mu(\mathbf{x}) + \frac{a}{2}\Delta(\mathbf{x}) \ ,$$

where $\mu(\mathbf{x}) = \frac{1}{2}\{E(Y|A = 1, \mathbf{X} = \mathbf{x}) + E(Y|A = -1, \mathbf{X} = \mathbf{x})\}$, which is a simple average of conditional outcomes under two treatments for a given set of covariate values. We refer to $\mu(\mathbf{X})$ as the main effect of $\mathbf{X}$ for convenience. Please note that this definition differs from the commonly used notion of "main effect" in the ITR literature, which typically refers to the conditional mean outcome under control ($E(Y|A = 0, \mathbf{X} = \mathbf{x})$).

2.2. *Reluctant Additive Individualized Treatment Rules.* In this section, we describe how we model and estimate ITRs by minimizing a loss function motivated by (2.3). Though linear ITRs are more interpretable than black-box ITRs, they may be insufficient in accurately describing complex HTEs. We aim to strike a balance between interpretability and complexity by extending linear ITRs via sparse generalized additive models (GAMs, Hastie and Tibshirani (1990)), which allows each variable to impact the outcome in a nonlinear and additive fashion, but omits interactions between components in $\mathbf{X}$. We adopt the "reluctant GAM" idea (Tay and Tibshirani, 2020) of fitting sparse GAMs, wherein linear approximations of $\Delta(\mathbf{x})$ are favored over nonlinear ones unless otherwise warranted by the data. As a guiding principle, we prefer an ITR to only contain effects that are linear in the original set of variables: non-linearities are only included thereafter if they significantly improve performance (Tay and Tibshirani, 2020). Since we focus on estimating treatment rules, we coin our approach as Reluctant Additive Individualized Treatment Rules (RAITR). An outline of the RAITR procedure is provided in Algorithm 1. Our approach differs from the approach of Tay and Tibshirani (2020) in several key ways, some of which are due to the particularities in estimating ITRs, and others are general modifications that we have found to improve performance and mitigate the effects of over-fitting arising from the procedure's sequential nature.

We propose to approximate $\Delta(\mathbf{X})$ with an additive model, i.e.

$$(2.4) \qquad \Delta(\mathbf{X}) \approx f(\mathbf{X}) \equiv \sum_{j=1}^{p} f_j(X_j),$$

where $f_j$ is a function of only the $j$th entry, $X_j$, of $\mathbf{X}$. In the following, we will introduce a computationally efficient and effective procedure for estimating the additive components $f_j$ that will comprise our estimated ITR.

Many frameworks have been proposed to model ITRs with the primary focus on linear ITRs including outcome-weighted learning (Zhao et al., 2012), Q-learning (Qian and Murphy, 2011), weighted learning (Chen et al., 2017), and dynamic weighted ordinary least squares (Wallace and Moodie, 2015). We propose a general framework that can enhance any least square loss-based linear ITR learning method (Qian and Murphy, 2011; Chen et al., 2017; Wallace and Moodie, 2015) to learn interpretable nonlinear ITRs. As an illustrative example, we focus on extending the weighted learning framework of Chen et al. (2017) to learn interpretable nonlinear ITRs. The loss function for learning optimal ITR ($f_{opt}(\mathbf{X})$) under weighted learning is defined as follows:

$$(2.5) \qquad f_{opt}(\mathbf{X}) = \mathrm{argmin}_f E\left[\frac{\left\{Y - m(\mathbf{X}) - \frac{A}{2}f(\mathbf{X})\right\}^2}{\pi(\mathbf{X}, A)}\right]$$

where $m(\mathbf{X})$ is any function of $\mathbf{X}$ only and $\pi(\mathbf{X}, A)$ is the propensity score. The validity of (2.5) is not impacted by the choice of $m(\mathbf{X})$ when $\pi(\mathbf{X}, A)$ is correctly specified. However, using $m(\mathbf{X}) = \mu(\mathbf{X})$ can protect against the misspecification of $\pi(\mathbf{X}, A)$. For this reason, we use this choice throughout our paper and generally advocate for use of flexible models to estimate this function. The inclusion of the inverse of propensity score is to eliminate measured confounding in observational studies. In particular, this weighting framework can be considered as the extension of inverse propensity score weighted regression for average treatment effect estimation to ITRs estimation.

Since we are not interested in the interpretation of the main effects and the propensity score, we can use flexible machine learning techniques to estimate them so that the model misspecification issues can be avoided. However, using flexible models for $m(\mathbf{X})$ and $\pi(\mathbf{X}, A)$ can also lead to over-fitting issue and bias the estimation of ITR. To mitigate this issue, we use the cross-fitting procedure (Chernozhukov et al., 2018) together with flexible models to estimate these nuisance parameters. For the rest of the section, we assume that we have access to an estimate of $m(\mathbf{X})$ and $\pi(\mathbf{X}, A)$ that can be plugged into (2.5), and we defer from describing cross-fitting version of our algorithm to Section 2.3.

Consider the case with $n$ independent observations $(Y_i, A_i, \mathbf{X}_i)_{i=1}^n$. The first step of our proposed procedure is to construct an initial sparse linear approximation of $\Delta(\mathbf{X})$ using weighted learning (Tian et al., 2014; Chen et al., 2017). This involves fitting a linear model for $\Delta(\mathbf{X})$ with a regularization term. For simplicity of presentation, we consider the LASSO (Tibshirani, 1996), but other penalties can also be used,

$$(2.6) \qquad \hat{\boldsymbol{\beta}}_1 = \operatorname{argmin}_{\boldsymbol{\beta}} \sum_{i=1}^n \hat{\pi}(\mathbf{X}_i, A_i)^{-1} \left\{ Y_i - \hat{m}(\mathbf{X}_i) - \frac{A_i}{2} \boldsymbol{\beta}^T \mathbf{X}_i \right\}^2 + \lambda_1 \sum_{j=1}^p |\beta_j| \, .$$

This results in coefficients $\hat{\boldsymbol{\beta}}_1$, which yield linear ITRs. We use $K$-fold cross validation to choose the tuning parameter in (2.6) with the weighted mean-squared error as the criterion. In particular, we use the corresponding inverse propensity score to weight the error contribution of each subject. Thus, the tuning parameter is selected to optimize accuracy in estimating the linear portion of the interaction function $\Delta(\mathbf{X})$.

The next step is to construct residuals from the linear ITRs and the main effects, $r_i = Y_i - \hat{m}(\mathbf{X}_i) - A_i \hat{\boldsymbol{\beta}}_1^T \mathbf{X}_i / 2$, for $i = 1, \ldots n$. Thus, if the additive approximation (2.4) is well-posited, the residual $r_i$ will contain only the error term and any nonlinear components of $\Delta(\mathbf{X})$. Using these residuals as a working response, for each covariate $X_j$ ($j = 1, \ldots p$) in $\mathbf{X}$, we fit an inverse propensity weighted penalized smoothing spline model (Wahba, 2006) to estimate the additive nonlinear contribution of each covariate $j$ to $\Delta(\mathbf{X})$. We then construct a design matrix of the fitted values from each of these covariate-specific models to be used as additional features in a final estimate of $\Delta(\mathbf{X})$. Constructing the nonlinear features one variable at a time ensures computational efficiency, while still maintaining an additive model (i.e., no between-variable interactions). In contrast to Tay and Tibshirani (2020), which used a fixed value for degrees of freedom, we use penalized splines with a data-adaptive value for degrees of freedom (Krivobokova et al., 2008). While using fixed degrees of freedom as in Tay and Tibshirani (2020) allows one to straightforwardly bound the overall degrees of freedom used, we have found that nonlinear terms for different covariates can be either under- or over-smoothed if there is a wide degree of variation in the variability of the different additive terms. We have found this problem to be well-mitigated by using penalized splines.

Let $\mathbf{G}_i$ denote the vector of nonlinear effects (i.e. the fitted values from the covariate-specific penalized splines) for the $i$th observation, where the $j$th entry of $\mathbf{G}_i$ is given by $g_j(X_{ij})$, where $g_j(\cdot)$ is the spline function fit by regressing $r_i$ on the $j$th covariate $X_{ij}$. We define the combined coefficients for the linear and nonlinear components as $\boldsymbol{\beta}^T = (\boldsymbol{\beta}_{lin}^T, \boldsymbol{\beta}_{non}^T)$ and their estimates as $\hat{\boldsymbol{\beta}}^T = (\hat{\boldsymbol{\beta}}_{lin}^T, \hat{\boldsymbol{\beta}}_{non}^T)$. Finally, in order to select the final model, we perform a penalized regression of $Y_i - \hat{m}(\mathbf{X}_i)$ on $\mathbf{X}_i$ and $\mathbf{G}_i$,

$$(2.7) \qquad \hat{\boldsymbol{\beta}} = \operatorname{argmin}_{\boldsymbol{\beta}} \Bigg( \sum_{i=1}^n \hat{\pi}(\mathbf{X}_i, A_i)^{-1} \left\{ Y_i - \hat{m}(\mathbf{X}_i) - \frac{A_i}{2} \left( \boldsymbol{\beta}_{lin}^T \mathbf{X}_i + \boldsymbol{\beta}_{non}^T \mathbf{G}_i \right) \right\}^2$$
$$+ \lambda_2 \sum_{j=1}^p \left( |\beta_{lin,j}| + \gamma_j |\beta_{non,j}| \right) \Bigg),$$

where $\gamma_j$ is a data-driven penalty factor for the $j$th nonlinear term. In particular, we set $\gamma_j = \min(\sqrt{p}, 1 + s_j^{-1})$ with $s_j$ defined as the standard deviation of the $j$-th nonlinear term across all observations. Such a penalty factor allows us to penalize nonlinear terms based on their complexity, i.e. the penalty factor for the $j$th nonlinear term is smaller for covariates whose estimated nonlinear terms have large variation and the penalty is larger for terms with small variation. Since $\gamma_j$ is lower bounded by 1, it guarantees the penalization for nonlinear terms is heavier than the corresponding linear term, which agrees with our proposal to have a parsimonious rule. Furthermore, $\gamma_j$ is upper bounded by $\sqrt{p}$ to prevent the overall penalty from being dominated by extreme penalty factors. The selection of the penalization term $\lambda_2$ will be discussed in more detail in Section 2.4, however it requires great care as to not over-fit the data, since the criterion (2.7) contains several terms ($\hat{m}$ and $\mathbf{G}$) estimated from the data. Given a well-chosen value $\lambda_2$, the final estimate for $\Delta(\mathbf{X})$ is expressed as

$$(2.8) \qquad \hat{f}_{\lambda_2}(\mathbf{X}) = \widehat{\boldsymbol{\beta}}_{lin}\mathbf{X} + \widehat{\boldsymbol{\beta}}_{non}\mathbf{G},$$

where inclusion of $\lambda_2$ in the notation indicates the dependence of the estimated function on the final tuning parameter. The final treatment rule is then $\hat{d}(\mathbf{X}) = \text{sign}\left\{\hat{f}_{\lambda_2}(\mathbf{X})\right\}$. We note that both (2.6) and (2.7) can be optimized with existing software for the LASSO, such as `glmnet` and thus each step is computationally simple and expedient.

---

**Algorithm 1** *Reluctant GAMs for Interpretable Nonlinear ITR*

---

**Input:** $(\mathbf{X}_i, Y_i, A_i)$, $i = 1, \ldots, n$: covariate $\mathbf{X}$ (with dimension $p$), outcome $Y$, treatment $A$, and penalty terms $\lambda_1, \lambda_2 \geq 0$.

**Process:**

1: Fit the weighted learning with least square loss and linear rule:

$$\widehat{\boldsymbol{\beta}}_1 = \text{argmin}_{\boldsymbol{\beta}} \sum_{i=1}^{n} \hat{\pi}(\mathbf{X}_i, A_i)^{-1} \left\{ Y_i - \hat{m}(\mathbf{X}_i) - \frac{A_i}{2}\boldsymbol{\beta}^T\mathbf{X}_i \right\}^2 + \lambda_1 \sum_{j=1}^{p} |\beta_j|,$$

and compute residuals $r_i = Y_i - \hat{m}(\mathbf{X}_i) - (A_i/2)\widehat{\boldsymbol{\beta}}_1^T\mathbf{X}_i, i = 1, \ldots n$.

2: For each covariate $j \in \{1, \ldots, p\}$, fit a weighted ($\hat{\pi}(\mathbf{X}_i, A_i)^{-1}$ as the weight) penalized smoothing spline (with data-dependent degrees of freedom) of $r_i$ on $g_j(X_{ij})$.

3: Fit penalized regression with LASSO penalty of $A$ on $\mathbf{X}$ and $\mathbf{G}$:

$$\widehat{\boldsymbol{\beta}} = \text{argmin}_{\boldsymbol{\beta}} \sum_{i=1}^{n} \hat{\pi}(\mathbf{X}_i, A_i)^{-1} \left\{ Y_i - \hat{m}(\mathbf{X}_i) - \frac{A_i}{2}\left(\boldsymbol{\beta}_{lin}^T\mathbf{X}_i + \boldsymbol{\beta}_{non}^T\mathbf{G}_i\right) \right\}^2$$

$$+ \lambda_2 \sum_{j=1}^{p} (|\boldsymbol{\beta}_{lin,j}| + \gamma_j|\boldsymbol{\beta}_{non,j}|),$$

with penalty $\gamma_j = \min(\sqrt{p}, 1 + s_j^{-1})$.

**Output:** $\hat{f}_{\lambda_2}(\mathbf{X}_i) = \widehat{\boldsymbol{\beta}}_{lin}\mathbf{X}_i + \widehat{\boldsymbol{\beta}}_{non}\mathbf{G}_i$ for $i = 1, \ldots, n$.

---

The proposed procedure has multiple aspects for encouraging parsimonious ITRs and enhancing their interpretability. Following the reluctant additive principle, our method promotes preference for linear terms in two ways: 1.) nonlinear terms are fit using the residuals after removing the linear terms; 2.) nonlinear terms are penalized based on their complexity, therefore they must contribute substantially to the ITR to survive the final model selection process in Step 3. Furthermore, traditional GAMs allow multiple nonlinear features representing the nonlinear effect of variable $j$. In contrast, our construction in Step 2 combines them as one feature. This ensures that after variable selection, one variable can contribute at most two terms to the final ITRs: one for the linear effect and the other for the nonlinear effect. This preference for linear terms combined with the simple structure as shown in (2.8) allows for complexity, while still maintaining an easy-to-understand interpretation, which is a key advantage of our method compared to existing nonlinear ITR learning methods. Now that we have shown the main idea of our approach, we next provide implementation details that are critical to making RAITR work effectively: i.e., estimation of $m(\mathbf{X})$ and $\pi(\mathbf{X}, A)$, and the tuning parameter ($\lambda_2$) selection.

2.3. *Nuisance parameter estimation with cross-fitting.* Since we are focused on identifying individualized treatment rules, the parameters in the main effects $m(\mathbf{X})$ and propensity score $\pi(\mathbf{X}, A)$ models may be considered nuisance parameters. The work by Pan and Zhao (2021) shows that using flexible models for estimating $m(\mathbf{X})$ and $\pi(\mathbf{X}, A)$ is still useful in the sense that well-estimated main effects and propensity scores can lead to improved robustness and efficiency of the ITR estimator. However, in practice, we have seen that using a model that is too flexible for the nuisance parameter can introduce bias. In order to use a flexible model while simultaneously preventing over-fitting, we use cross-fitting (Chernozhukov et al., 2018) with a gradient boosting model (Hastie et al., 2009) to estimate the main effects model. Using complex models for the nuisance parameters does not conflict with our goal of creating interpretable ITRs, since this does not affect the interpretability of the final treatment rules. Once the main effects and propensity score estimates are specified, we implement the augmentation procedure with our approach as described in Section 2.2. The estimation of nuisance parameters is performed as follows:

1) Take a $K$-fold random partition $(I_k)_{k=1}^K$ of observation indices $[n] = 1, ..., n$ such that the size of each fold $I_k$ is $n_K = n/K$. Also, for each $k \in [K] = 1, ..., K$, define the complement set $I_k^c := \{1, ..., n\} \backslash I_k$.
2) For each $k \in K$, using only data in using $I_k^c$, we construct the propensity score model, and the main effect models. We denote these models as $\hat{\pi}^{I_k^c}(\mathbf{X}, A)$, and $\hat{m}^{I_k^c}(\mathbf{X})$ respectively.

Since the construction of the propensity score model and main effect model uses the out-of-fold samples, cross-fitting prevents the model over-fitting for these nuisance parameters. Though cross-fitting to avoid overfitting of nuisance models is important, it is worth noting that the objective for the two nuisance models differs: optimizing fit (minimizing prediction error) in the main effect model can reduce variance and improve ITRs, whereas optimizing fit in the propensity score may lead to instability: the goal of this model is to achieve covariate balance between treatment groups, and ideally minimizing covariate imbalance should be the objective of this model fit.

Once we estimate the nuisance parameters, to implement RAITR, we simply replace $\hat{\pi}(\mathbf{X}, A)$ and $\hat{m}(\mathbf{X})$ in Algorithm 1 with $\hat{\pi}^{I_k^c}(\mathbf{X}, A)$, and $\hat{m}^{I_k^c}(\mathbf{X})$ respectively. Adapting Algorithm 1 to use with estimators arising from out-of-fold data requires care, therefore we explicitly show an algorithm (Algorithm 2) in the supplementary file.

Now that we have described fitting the RAITR model, including the nuisance parameters, the final requirement is selecting the tuning parameter, $\lambda_2$, for the penalized regression of the final model. Next, we specify the criteria for selecting this parameter.

2.4. *Model Selection Criterion.* A natural choice for selecting $\lambda_2$ (and the corresponding ITR) is to perform $K$-fold cross validation of the LASSO model in (2.7). However, an extra stage of cross validation would increase required computation. Moreover, as we show in simulations and data example, cross validation tends to over-select nonlinear effects. Lastly, conducting cross validation for the final model can attenuate the benefit of cross-fitting i.e., the estimation error of nuisance parameters and parameters of interest in the ITR would no longer be orthogonal unless complex nested cross validation strategy is used. Hence, we consider an alternative approach for selecting $\lambda_2$ by the Concordance Information Criterion (CIC; Shi et al. (2021)). From our experimentation, CIC performs favorably to cross-validation to control bias in the final ITRs of RAITR. As an added feature, the CIC does not require splitting the data into folds, thereby reducing computational load. The CIC is denoted

$$\text{CIC}(\hat{f}_\lambda) = nC(\hat{f}_\lambda) - \kappa_n DF(\hat{f}_\lambda),$$

where $C(\hat{f}_\lambda)$ is the concordance function for the ITR ($\text{sign}(\hat{f}_\lambda)$) governed by a tuning parameter $\lambda$ that drives the complexity of the ITR, $\kappa_n$ is a constant depending on $n$ and/or $p$, and $DF(\hat{f}_\lambda)$ is the degrees of freedom of the corresponding ITR. In our setup, we use $\lambda = \lambda_2$. Choices for the value of $\kappa_n$ and $DF(\hat{f}_\lambda)$ may affect the performance of CIC and optimal choices are still open research problems, however, we aim to provide recommendations based on numerical results. The CIC is composed of two terms, the first term is the concordance function, and the second is a penalty for ITR complexity, similar to BIC. The concordance function is defined as

$$C(\hat{f}_\lambda) = E[\{Y_i(1) - Y_i(-1)\} - \{Y_j(1) - Y_j(-1)\}] \times \mathbb{1}\{\hat{f}_\lambda(\mathbf{X}_i) > \hat{f}_\lambda(\mathbf{X}_j)\}.$$

Let $A = -1$ denote the reference treatment (either control, placebo, or active comparator treatment) and $A = 1$ denote the treatment of interest, the concordance is a measure of the degree to which patients who are more likely to be assigned to the treatment of interest also have larger individual treatment effects (relative to the reference treatment $A = -1$). The empirical estimator of the concordance is given by,

(2.9)
$$\hat{C}(\hat{f}_\lambda) = \frac{1}{n(n-1)} \sum_{i \neq j} \left[ \frac{\{Y_i - E(Y_i | \mathbf{X}_i = \mathbf{x}_i, A_i = -1)\}\{0.5A_i + 0.5 - \pi(\mathbf{x}_i, A_i = 1)\}}{\pi(\mathbf{x}_i, A_i = a_i)} \right.$$

$$\left. - \frac{\{Y_j - E(Y_j | \mathbf{X}_i = \mathbf{x}_i, A_i = -1)\}\{0.5A_j + 0.5 - \pi(\mathbf{x}_j, A_i = 1)\}}{\pi(\mathbf{x}_j, A_j = a_j)} \right] \times \mathbb{1}\left\{\hat{f}_\lambda(\mathbf{x}_i) > \hat{f}_\lambda(\mathbf{x}_j)\right\},$$

We also consider the doubly-robust estimator,

(2.10)
$$\hat{C}_{\text{DR}}(\hat{f}_\lambda) = \frac{1}{n(n-1)} \sum_{i \neq j} \left[ \frac{\{Y_i - E(Y_i | \mathbf{X}_i = \mathbf{x}_i, A_i = -1)\}\{0.5A_i + 0.5 - \pi(\mathbf{x}_i, A_i = 1)\}(0.5A_j + 0.5)}{\pi(\mathbf{x}_i, A_i = a_i)\pi(\mathbf{x}_j, A_j = 1)} \right.$$

$$\left. - \frac{\{Y_j - E(Y_j | \mathbf{X}_j = \mathbf{x}_j, A_j = -1)\}\{0.5A_j + 0.5 - \pi(\mathbf{x}_j, A_i = 1)\}(0.5A_i + 0.5)}{\pi(\mathbf{x}_i, A_j = a_j)\pi(\mathbf{x}_i, A_i = 1)} \right]$$

$$\times \mathbb{1}\left\{\hat{f}_\lambda(\mathbf{x}_i) > \hat{f}_\lambda(\mathbf{x}_j)\right\}.$$

$\hat{C}_{\text{DR}}(\hat{f}_\lambda)$ is doubly robust in that it is a consistent estimator of $C(\hat{f}_\lambda)$ when either the models for $\pi(\mathbf{x}, A)$ or $E(Y_i | \mathbf{X} = \mathbf{x}, A_i = -1)$ is correctly specified. Although Fan et al. (2017) and Shi et al. (2021) studied the theoretical properties of these estimators using $\kappa_n$ of a certain order, we consider the numerical performance of different choices of $\kappa_n$. Particularly, we show two possible choices $\kappa_n = \log n$ or $\kappa_n = 2$, which were advocated by Shi et al. (2021) to mimic AIC and BIC for likelihood-based model selection. In the context of RAITR, we maximize the CIC to select the tuning parameter in our final model. Note that we can plug in the cross-fitted version of the main effect model ($m^{I_k^c}$) and propensity score model ($\pi^{I_k^c}$) into (2.9 or 2.10) such that the resulting criterion may benefit from the nuisance parameter estimation described in Section 2.3. Obtaining a precise estimate of $DF(f_\lambda)$ is challenging in our setting and is an area of open work. We use a simple approximation that we have found is effective in practice. The choice of $DF(f_\lambda)$ is $\sum_{j=1}^{p} \mathbb{1}\{\hat{\beta}_{lin,j} \neq 0\} + \sum_{j=1}^{p} \mathbb{1}\{\hat{\beta}_{non,j} \neq 0\}\widehat{\text{DF}}(g_j)$, where $\widehat{\text{DF}}(g_j)$ is the estimated degrees of freedom for the estimated smoothing spline for the $j$th variable.

In our proposal, we use $\lambda = \lambda_2$ and do not select $\lambda_1$ with the CIC metric, as $\lambda_1$ is set in the first stage. We do not choose $\lambda_1$ with CIC because we do not use the linear terms by themselves as our final treatment rule. Instead, our choice of $\lambda_1$ emphasizes accuracy in predicting the linear components of the interactions, whereas our choice of $\lambda_2$ emphasizes performance of the estimated ITRs resulting from our entire proposed estimation procedure.

**3. Simulation.** In this section, we study the performance of our method and existing methods in a number of settings. In each setting, we simulate $n$ i.i.d samples $(Y_i, \mathbf{X}_i, A_i)$. Each $X_j$ $(j = 1 \ldots p)$ is distributed as $X_j \sim \text{Unif}(-2, 2)$, where 'Unif' represents the continuous uniform distribution. The treatment $A$ follows a binary distribution ($-1$ or $1$), where $P(A = 1 | \mathbf{X}) = \exp(\mathbf{X}^T \beta) / \{1 + \exp(\mathbf{X}^T \beta)\}$, with $\beta = (1, -1, 0.5, 0, 0, \ldots, 0)^T$. The response is normally distributed as $N(m(\mathbf{X}) + (A/2) \times \Delta(\mathbf{X}), 4)$. For these simulations, $m(\mathbf{X})$ is given by the nonlinear function $m(\mathbf{X}) = -c \sum_{j=1}^{5} \{X_j + (2/3)(2X_j^2 - 1)\}$. We adjust the value of $c$ to adjust the relative effect size of $\Delta(\mathbf{X})$ to the remaining components in the response function. For each data-generating model we investigate two values of $c$, the smaller value indicates a "large" effect size and the larger value indicates a "small" effect size. We consider multiple models for $\Delta(\mathbf{X})$ with various degrees of complexity. The goal is to show that in cases where $\Delta(\mathbf{X})$ is relatively simple (e.g. linear), our approach does not sacrifice simplicity, however in complicated cases (e.g. highly nonlinear, polynomial, or tree-based models), RAITR is able to capture these complexities. The particular choices of $\Delta(\mathbf{X})$ are given as below:

1. Linear: $\Delta(\mathbf{X}) = \mathbf{X}^T \beta$, $\beta = (1, -1, 0.5, 1, -1, -0.5, 0, 0, \ldots, 0)^T$, $c = 0.1$ or $3$
2. Highly Nonlinear: $\Delta(\mathbf{X}) = X_1^3 + |X_3| \exp\{X_5\} + 5 \sin(2\pi X_7) + 5 \cos(2\pi X_6) - (X_4 + X_8)^2 + 3|X_2 + X_5|$, $c = 1$ or $8$
3. Tree: $\Delta(\mathbf{X}) = 3\{\mathbb{1}(X_1 + 0.5 > 0) \times \text{sign}(X_2 - 0.5)\} + 2.5\{\mathbb{1}(X_1 + 0.5 < 0) \times \text{sign}(X_4 - 0.5)\} + 0.5$, $c = 1$ or $5$
4. Polynomial: $\Delta(\mathbf{X}) = 0.2 + X_1^2 + X_2^2 - X_3^2 - X_4^2$, $c = 0.1$ or $2$

The optimal ITRs are additive in $\mathbf{X}$ in the linear and polynomial settings, but are not additive in the highly nonlinear and tree settings.

We train the ITRs with various sample sizes ($n = 250, 500, 1000$) and covariate dimension ($p = 100, 500, 1000$) combinations, then evaluate the ITR on a testing set of size $10,000$. We report two commonly used metrics for ITR performance evaluation: the value function of the estimated rule and the agreement between the estimated rule and the optimal rule. In particular, for a given rule $\hat{d}(\mathbf{X})$, we can directly estimate $E[Y(\hat{d}(\mathbf{X}))]$ and $E[\hat{d}(\mathbf{X}) = d^*(\mathbf{X})]$ by evaluating their empirical versions on the testing data.

We estimate ITRs using two versions of a strictly linear rule based on the weighted learning (Chen et al., 2017) with the least square loss (i.e., using only the first three steps of the algorithm in Supplement Section A to estimate ITRs), generalized random forests (GRF; Athey et al. (2019); Wager and Athey (2018)), sparse additive model for treatment effect-modifier selection (samTEMsel; Park et al. (2022)), and different variations of our method, RAITR. We did not include support vector machine-based nonlinear ITR learning methods such as Zhao et al. (2012); Zhou et al. (2017) in the comparison as they are not suited for the high dimensional covariate settings considered in this paper. Additional information about the simulation scenarios including computational load, covariate balancing, and a formal definition of "small" and "large" effect sizes can be found in Supplement Sections B.1, B.2, and B.3, respectively.

In Figures 1 to 4, we show results across 100 replicates in terms of the agreement. In Supplement Section B.3, we also display results in terms of the value function. For linear

ITRs and RAITR, we either select tuning parameters using cross-validation (CV) or the concordance information criterion (CIC). When CIC is used, we either set $\kappa_n = \log n$ or $\kappa_n = 2$ ($2\kappa$), and either use the standard estimate for concordance or the doubly-robust (DR) version. For other methods, the tuning parameters are selected by CV.

Regarding the linear setting, unsurprisingly, the linear ITR analyses perform best, but RAITR, particularly with tuning parameters selected using CIC, performs comparably to the linear rules. Using RAITR with final models selected by CV generally performs poorly, because the model over-selects nonlinear terms. The samTEMsel method generally performs relatively poorly compared to linear ITRs, suggesting a tendency for it to select overly complex ITRs when the underlying data structure is simple. Similarly, GRF struggles due to the linear ITR structure and is particularly poor in high-dimensional settings, as it does not have an explicit variable selection component.

In the highly nonlinear setting, RAITR with CIC consistently outperforms all other methods, particularly with a large effect size. Due to the highly complex and non-additive nature of the ITR in this setting, there appears to be an upper bound on how well RAITR can perform. Despite this, in the "large" effect setting, RAITR is desirable compared to the other nonlinear ITRs. In the "small effect" setting, no method except RAITR (with a large enough sample size) does better than the naive "always treat" rule.

In the tree setting, GRF outperforms all other methods for some scenarios but performs poorly when the dimension is large, again due to its lack of an explicit variable selection component. Using samTEMsel or RAITR with CIC consistently performs well. In the polynomial setting, RAITR and samTEMsel perform much better than all other methods, with RAITR performing the best when the sample size is large enough to support the augmentation procedure.

In general, RAITR with CIC regularly performs at least as well, and many times, better than the other approaches considered. A caveat is that the augmentation procedure tends to require a large sample size to perform well. However, even in the settings with smaller sample size, RAITR performs comparably to other approaches, and in the settings with large sample size often performs much better. In many cases, when $\kappa_n = \log n$, RAITR performs better on average than when $\kappa_n = 2$, however it appears that the $2\kappa$ version can better protect against poor performance. Therefore, the choice of $\kappa_n$ should be made relative to the application of interest. In these simulations, RAITR with doubly-robust CIC performed comparably to the standard version. Additionally, RAITR performs well in the "small effect" setting, particularly when the sample size is large enough to support augmentation. Finally, increasing dimensionality can cause deterioration in performance for other methods, such as GRF, samTEMsel, and linear rules, but RAITR appears to be more robust to this problem. Notably, the results for the linear ITR and highly nonlinear ITR settings jointly emphasize RAITR's ability to balance performance and return either simple ITRs or sufficiently complex ITRs when necessary. Note that, Shi et al. (2021) also proposed the value information criterion (VIC) for ITR model selection, as well as CIC. In their study, Shi et al. (2021) observed that CIC performed better than VIC empirically. We also observed a similar phenomenon in the early development of our method. As such, we focus on comparing the results of RAITR by using either CIC or the commonly used method of CV as the model selection criterion.

As a final point of emphasis, a key reason for RAITR to yield a highly interpretable result is that each covariate can only contribute at most two terms to the final ITR. This is demonstrated in Figure 5. For demonstration, we generate data using a similar model to the highly nonlinear setting as described before, except now we have $\Delta(\mathbf{X}) = (1/2)\cos(2X_1)(2X_1)$, $n$=1000, and $p = 10$. We show the contribution of $X_1$ to the resulting ITR from one repetition of RAITR (CIC, $2\kappa$, DR). On the left, we plot the covariate versus $\hat{f}(\mathbf{X}) = \widehat{\boldsymbol{\beta}}_{lin}\mathbf{X} + \widehat{\boldsymbol{\beta}}_{non}\mathbf{G}$, as well as the true ITR. On the top-right and bottom-right, we show only the linear and nonlinear contribution, respectively. The figure demonstrates that RAITR allows us to visualize the contribution of a covariate to an ITR in a simple manner, which results in a higher level of interpretability and can be quite useful in practice.

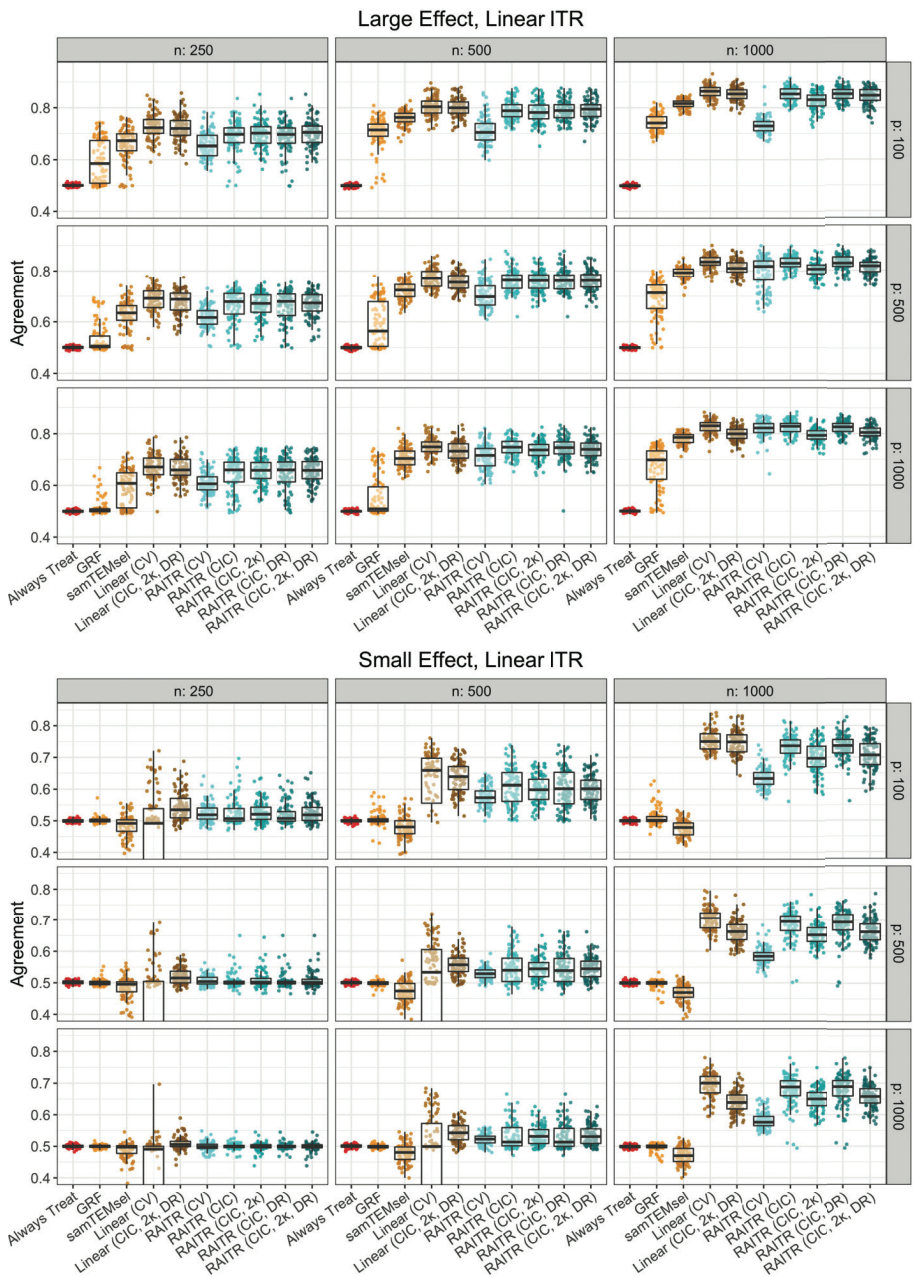

Fig 1: Top: results for agreement for large effect size, linear data-generating model for $n = 250,\ 500,\ 1000$ and $p = 100,\ 500,\ 1000$ for RAITR, linear ITRs, GRF, and samTEMsel. Bottom: as for top but small effect size. The vertical axis limits of the bottom plot have been modified for clarity.

**4. Analysis of GDSC Data.** The Genomics of Drug Sensitivity in Cancer (GDSC) study measures the response of $1,018$ human cancer cell lines to $265$ anti-cancer drugs (Yang

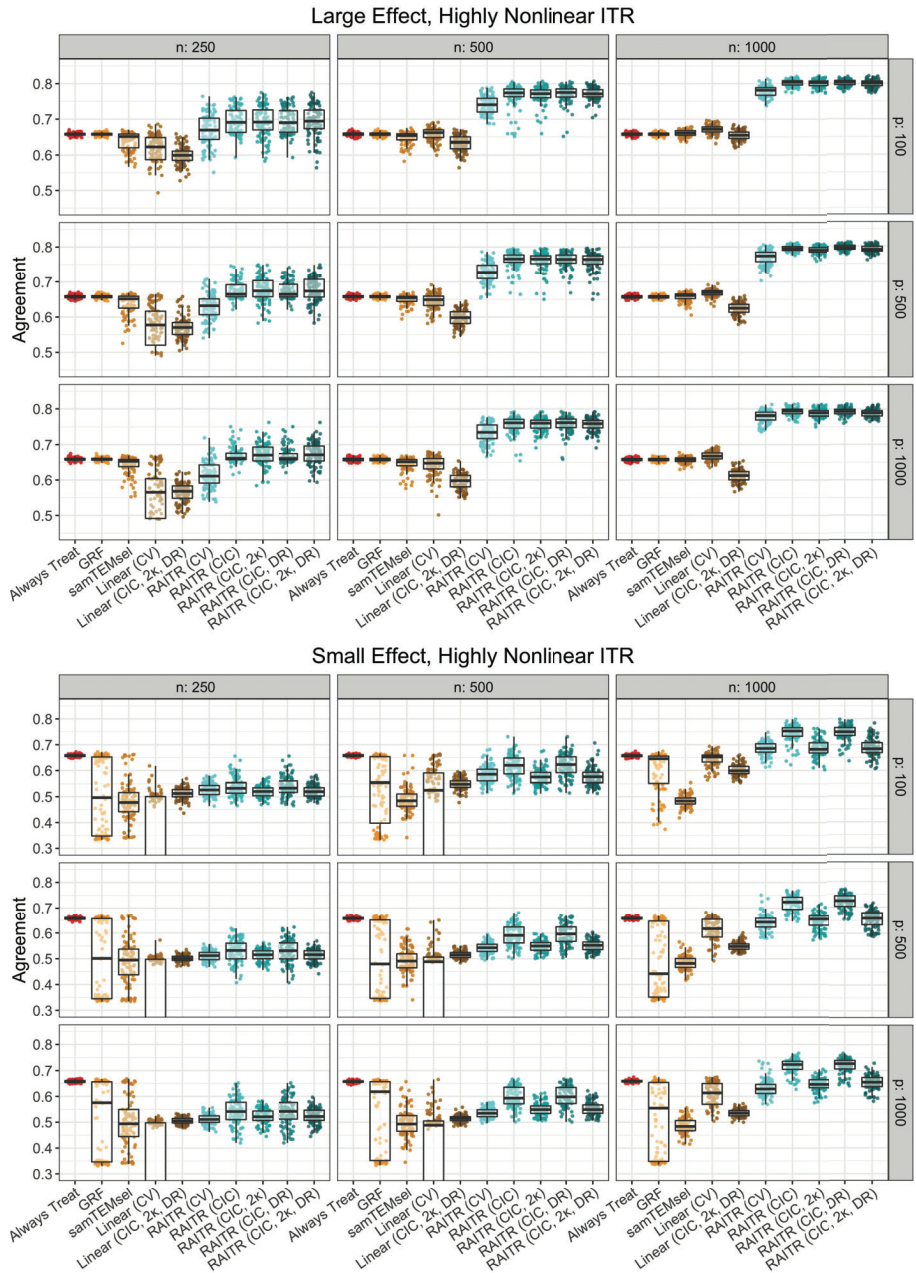

Fig 2: Top: results for agreement for large effect size, highly nonlinear data-generating model for $n = 250,\ 500,\ 1000$ and $p = 100,\ 500,\ 1000$ for RAITR, linear ITRs, GRF, and samTEMsel. Bottom: as for top but small effect size. The vertical axis limits of the bottom plot have been modified for clarity.

et al., 2012). In particular, it contains detailed genetic information, including gene expression and chromosomal copy number, pharmacological information and treatment responses on human cancer cell lines. As noted by Iorio et al. (2016), since cancer cell lines are replicable, we may test multiple drugs on the same cell line in order to compare drug sensitivity to a response of interest. This makes the GDSC data and similar datasets, such as Cancer Cell

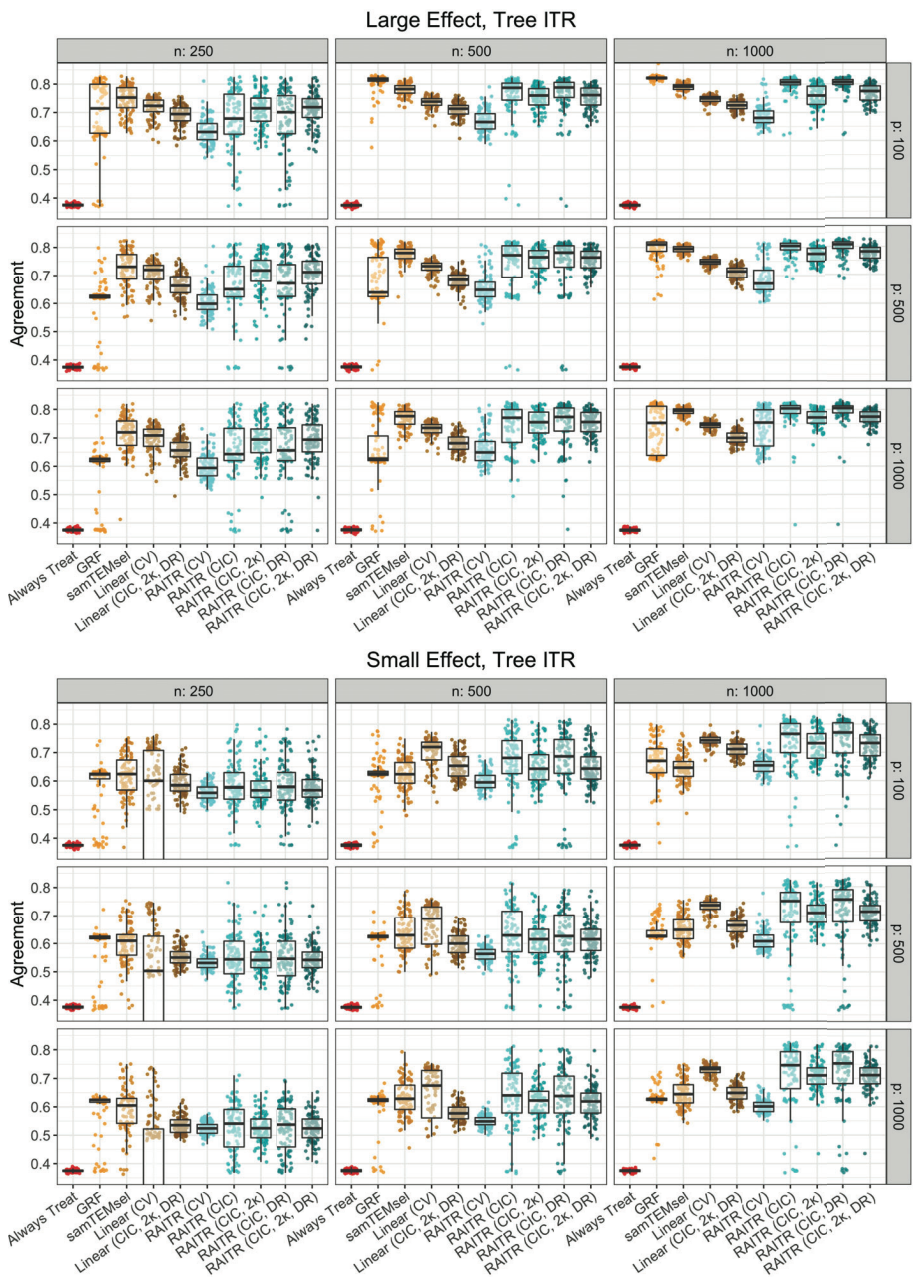

Fig 3: Top: results for agreement for large effect size, tree data-generating model for $n = 250,\ 500,\ 1000$ and $p = 100,\ 500,\ 1000$ for RAITR, linear ITRs, GRF, and samTEMsel. Bottom: as for top but small effect size.

Line Encyclopedia (Ghandi et al., 2019), a particularly interesting resource for benchmarking the performance of different ITRs.

In this paper, we focus on the outcome $IC_{50}$, which is the drug concentration that reduces viability by 50%. This measurement was collected on each of the $1,018$ cancer cell lines. A smaller value for $IC_{50}$ means the drug is more effective. Hence, to be consistent with the notation of a large value is desirable, we use $-\log(IC_{50})$ the log transformation is done following the procedure used by Yang et al. (2012) as the outcome. We use the gene expression data (17, 419 genes in total) of each cancer line as the predictors, and our aim is to recommend the drug that is more effective between a pair of drugs. Gene expression was chosen because it was the most predictive data type for drug sensitivity in the GSDC study (Iorio et al., 2016).

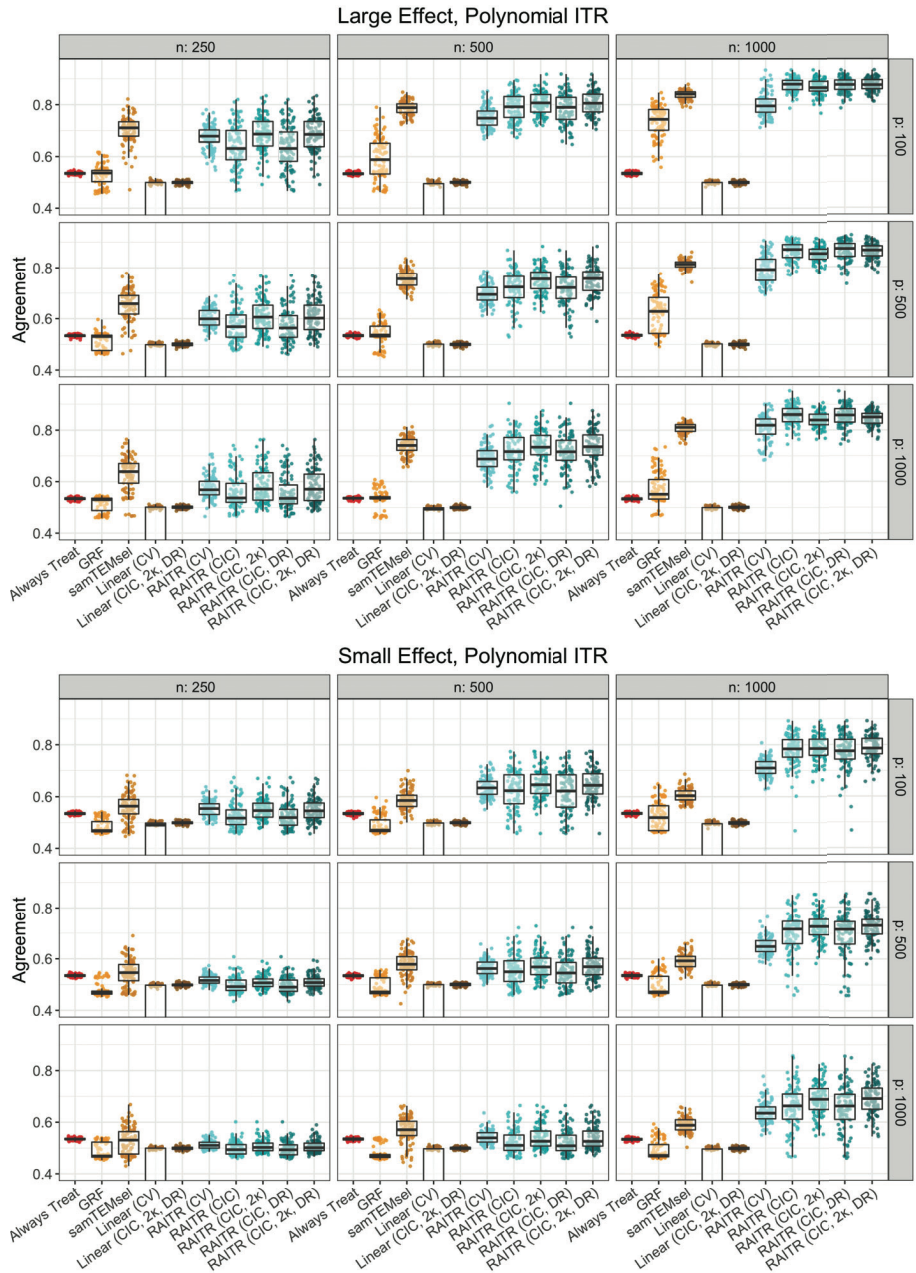

Fig 4: Top: results for agreement for large effect size, polynomial data-generating model for $n = 250,\ 500,\ 1000$ and $p = 100,\ 500,\ 1000$ for RAITR, linear ITRs, GRF, and samTEMsel. Bottom: as for top but small effect size.

The minimum and maximum of the mean $\log(\text{IC}_{50})$ of the considered drugs are $-5.59$ and $8.00$, respectively. There are 133 drugs with the mean $\log(\text{IC}_{50})$ less than 2. Among these drugs with relatively small $\log(\text{IC}_{50})$, there are 235 drug pairs with similar $\log(\text{IC}_{50})$ values among cancer cells: i.e., the empirical probability of one drug being the optimal treatment is greater than $0.35$ and less than $0.65$. Lastly, we select the 26 drug pairs out of 235 drug pairs that are most likely to have a nonlinear relationship between $\log(\text{IC}_{50})$ and the gene expression level. The complete results for all 26 pairs are given in Supplement Section C. Here, we show results from six drug pairs that are representative: (1) SN-38 and Vinblastine, (2) JQ1 and Navitoclax, (3) IGFR_3801 and CD532, (4) AZD2014 and IOX2, (5) Cytarabine and IMD-0354, and (6) Refametinib and AZD6738. These drug combinations were chosen because they gave a large variety in performance for the methods under consideration.

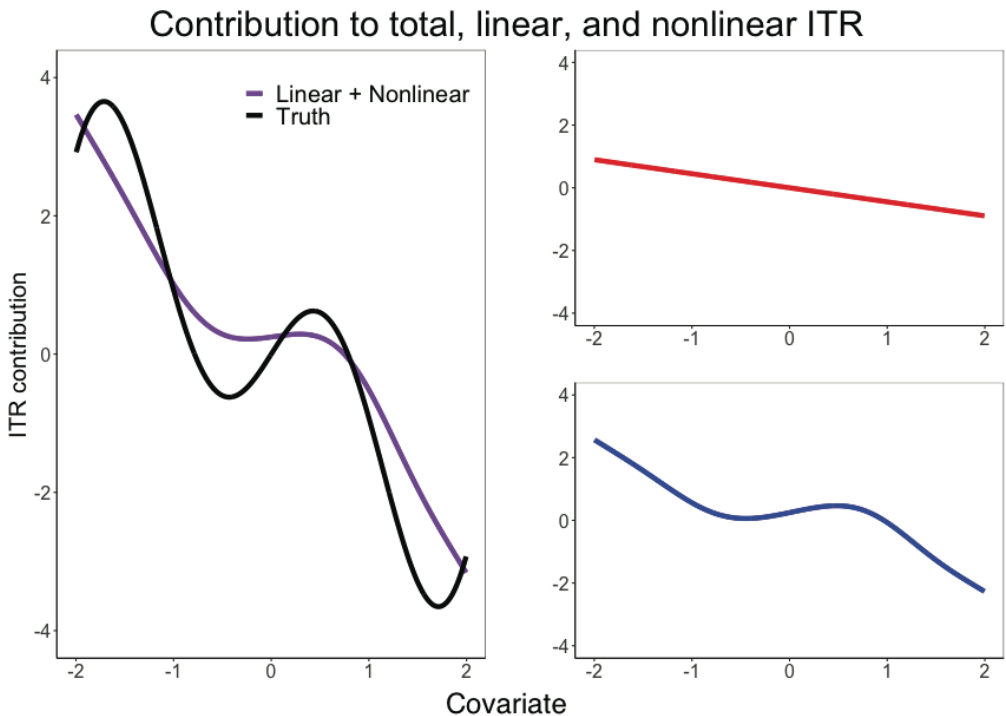

Fig 5: Example of resulting ITR from RAITR (CIC, $2\kappa$, DR) from one repetition of the highly nonlinear simulation setting (left), as well as contribution from the linear component (top-right) and nonlinear component (bottom-right).

We use the median absolute deviation (MAD) to filter out the gene expressions with low variability (MAD $< 1$), resulting in $1{,}486$ genes for the downstream analyses. The number of cells (sample size) varies for different drug pairs due to different degree of outcomes missing – a full table of number of cells for each of the 26 drug pairs is given in Table C.1 of the supplementary materials, but is approximately 900 cells for each drug pair.

We randomly split the data into training (two-thirds) and testing (one-third), then evaluate trained models on the testing data. We repeat this procedure 100 times. Even though we can observe the response for both treatments, we construct the training data to be more comparable to learning ITRs in a standard clinical setting. In particular, for each cell, we randomly select one treatment as the 'observed' treatment and the corresponding $-\log(\text{IC}_{\mathbf{50}})$ as the 'observed' outcome and discard the remaining treatment and outcome-related information for model training. On the testing data, we can directly calculate the empirical agreement and value function of the trained model on the testing data as we did in the simulation study. Typically, we would need to estimate the propensity score and/or outcome regression function for calculating the empirical agreement and the value function, but due to the special structure of the GDSC data, this is not required. Finally, we report the average agreement and value function over the 100 replications. A selection of the results with respect to the agreement is given in Table 1. The complete results with the agreement and the value function for all 26 drug pairs are given in Supplement Section C.

As seen in Table 1, RAITR, particularly RAITR (CIC, $2\kappa$, DR), performs well across all drug pairs, which indicates a high level of flexibility. In two out of the six reported drug pairs RAITR (CIC, $2\kappa$, DR) performs best, and in the remaining settings, still performs strongly. Each of the other methods performs well for particular drug pairs, but also performs poorly in other pairs. The high degree of flexibility, together with the aforementioned interpretability makes RAITR a desirable method for ITR identification.

TABLE 1

*Results from GDSC analysis for a selection of drug pairs. In one row, linear method is denoted L. to increase clarity. Results are given in terms of mean agreement across 100 replicates with the associated standard errors in parentheses. The highest performing method(s) for each drug is in bold.*

| | SN-38 and Vinblastine | JQ1 and Navitoclax | IGFR_3801 and CD532 | AZD2014 and IOX2 | Cytarabine and IMD-0354 | Refametinib and AZD6738 |
|---|---|---|---|---|---|---|
| Method | Mean agreement (standard deviation), 100 replicates | | | | | |
| Always Treat | 0.517 (0.022) | 0.593 (0.025) | 0.599 (0.025) | 0.536 (0.023) | 0.538 (0.025) | 0.496 (0.024) |
| GRF | 0.527 (0.030) | **0.659** (0.025) | 0.654 (0.029) | 0.534 (0.040) | 0.605 (0.034) | 0.681 (0.026) |
| samTEMsel | 0.562 (0.043) | 0.647 (0.025) | 0.656 (0.024) | 0.528 (0.041) | 0.576 (0.035) | 0.687 (0.027) |
| Linear (CV) | 0.627 (0.048) | 0.618 (0.042) | 0.653 (0.042) | 0.531 (0.041) | 0.603 (0.036) | **0.697** (0.026) |
| L. (CIC, $2\kappa$, DR) | **0.630** (0.034) | 0.621 (0.027) | 0.661 (0.026) | **0.551** (0.028) | 0.609 (0.029) | 0.696 (0.025) |
| RAITR (CV) | 0.613 (0.037) | 0.630 (0.033) | 0.655 (0.035) | 0.543 (0.037) | 0.599 (0.034) | 0.687 (0.029) |
| RAITR (CIC) | 0.600 (0.052) | 0.631 (0.042) | 0.645 (0.044) | 0.538 (0.036) | 0.590 (0.033) | 0.677 (0.037) |
| RAITR (CIC, $2\kappa$) | 0.627 (0.038) | 0.641 (0.032) | **0.667** (0.029) | 0.549 (0.031) | 0.612 (0.026) | **0.697** (0.026) |
| RAITR (CIC, DR) | 0.599 (0.054) | 0.631 (0.039) | 0.647 (0.038) | 0.539 (0.037) | 0.592 (0.032) | 0.680 (0.036) |
| RAITR (CIC, $2\kappa$, DR) | 0.628 (0.038) | 0.642 (0.031) | **0.667** (0.029) | 0.550 (0.030) | **0.613** (0.024) | 0.696 (0.026) |

**5. Discussion.** We propose a new method, RAITR, for learning flexible yet interpretable nonlinear ITRs based on a reluctant generalized additive model. RAITR is designed with high-dimensionality in mind and efficiently adapts to the smoothness of the underlying ITR. RAITR is parsimonious by supplying a simple modeling framework that only includes complex terms that are essential for the ITR. Even when complex, nonlinear terms are selected, our framework allows for a simple data-visualization tool in order to increase interpretability and collaboration. On the other hand, using the reluctant additive framework creates challenges, which we have addressed by combining cross-fitting with CIC, a modern information-based model selection criterion.

As is seen in both the data analysis and simulations, there are times when the linear ITR is enough for an accurate treatment decision. This is because, even if the underlying heterogeneous treatment effect is highly nonlinear, the resulting optimal ITR may be well-approximated by a linear ITR. As such, it is imperative to have highly flexible methods which perform well in simpler settings, such as RAITR. Despite the demonstrated advantages of RAITR, one key area of concern with respect to performance is that the augmentation procedure seems to require a relatively large sample size, particularly in high-dimensional settings, to perform well. For this reason, in extremely small sample size settings, we would recommend either using a simpler parametric model for augmentation or omitting the augmentation step (omitting would not impact bias). Furthermore, because our method is designed to "fall back" to a linear ITR when nonlinear terms are not warranted by the data, our approach is almost always competitive with linear ITR approaches. Lastly, regarding when it is appropriate to estimate an ITR, the general principle would also apply: if the sample size is simply too small it may not feasible to estimate an ITR using any method. If the dimensionality is extremely high, it is advisable to use a variable screening procedure to eliminate variables that do not have marginal interactions with the treatment.

Note that the current implementation of RAITR does not enforce the strong heredity property, which is the notion of including a nonlinear term in the model only when the corresponding linear term has also been included. Even though the chance of violation of strong heredity property with RAITR is low due to how the nonlinear terms are constructed, incorporating strong heredity would be an important future area of research. For example, one approach would be to use the penalty proposed in Bian et al. (2021) to ensure strong heredity. Moreover, even though RAITR is motivated by learning ITRs for observational studies, it is directly applicable to data from controlled experiments and randomized trials. Due to a commonly observed phenomenon of variance reduction when known propensity scores are estimated, the estimated propensity score would still be used in RAITR, since even a well-executed randomized trial could have finite sample covariate imbalance, and estimating average treatment effect has shown that it is beneficial to use the estimated (correctly-specified) propensity score even if the true treatment assignment mechanism is known (Hirano et al., 2003).

In this work, we focused on extending linear ITRs to nonlinear ITRs under the weighted learning framework, however, our method can be readily applicable to other ITR learning frameworks whose current implementations are focused on learning linear ITRs such as A-learning (Chen et al., 2017; Nie and Wager, 2021), dynamic weighted ordinary least squares (Wallace and Moodie, 2015), and more. Moreover, the RAITR approach was designed with continuous outcomes in mind, but the justification of the loss does not require the outcome to be continuous and could thus be applied to count or binary outcomes. However, a similar approach could be developed for other types of responses using the general class of loss functions explored in Chen et al. (2017) and Huling and Yu (2021). Other areas of future research include extending RAITR for multi-category treatments (Qi et al., 2020) or continuous treatments (Chen et al., 2016) and allowing for the inclusion of pairwise interaction effects via the reluctant interaction modeling framework (Yu et al., 2019).

**Acknowledgements.** The authors thank the Editor, the referees, and the Associate Editor for their insightful comments and suggestions. Guanhua Chen and Jared D. Huling are co-corresponding authors of the paper who jointly developed the method and supervised the project.

**Funding.** Maronge's effort towards this work was supported by NIH grant R01HL094786 and a University of Wisconsin - Madison Morse Fellowship. Chen's effort towards this work was partially supported by NSF grant DMS-2054346 and the University of Wisconsin School of Medicine and Public Health from the Wisconsin Partnership Program (Research Design Support: the Protocol Development, Informatics, and Biostatistics Module)

## SUPPLEMENTARY MATERIAL

Algorithm 2, additional simulation and real data analysis results, information on computation times, as well as the software for the proposed method in R, can be found in the Supplementary Material.

# REFERENCES

Athey, S., J. Tibshirani, and S. Wager (2019). Generalized random forests. *Annals of Statistics 47*(2), 1148–1178.

Bian, Z., E. E. Moodie, S. M. Shortreed, and S. Bhatnagar (2021). Variable selection in regression-based estimation of dynamic treatment regimes. *Biometrics*.

Chen, G., D. Zeng, and M. R. Kosorok (2016). Personalized dose finding using outcome weighted learning. *Journal of the American Statistical Association 111*(516), 1509–1521.

Chen, S., L. Tian, T. Cai, and M. Yu (2017). A general statistical framework for subgroup identification and comparative treatment scoring. *Biometrics 73*, 1199–1209.

Chernozhukov, V., D. Chetverikov, M. Demirer, E. Duflo, C. Hansen, W. Newey, and J. Robins (2018). Double/debiased machine learning for treatment and structural parameters. *The Econometrics Journal 21*(1), C1–C68.

Fan, C., W. Lu, R. Song, and Y. Zhou (2017). Concordance-assisted learning for estimating optimal individualized treatment regimes. *Journal of the Royal Statistical Society: Series B (Statistical Methodology) 79*(5), 1565–1582.

Ghandi, M., F. W. Huang, J. Jané-Valbuena, G. V. Kryukov, C. C. Lo, E. R. McDonald, J. Barretina, E. T. Gelfand, C. M. Bielski, H. Li, et al. (2019). Next-generation characterization of the cancer cell line encyclopedia. *Nature 569*(7757), 503–508.

Hastie, T., R. Tibshirani, and J. Friedman (2009). *The Elements of Statistical Learning*, Volume 1. Springer New York.

Hastie, T. J. and R. J. Tibshirani (1990). *Generalized additive models*, Volume 43. CRC press.

Hirano, K., G. W. Imbens, and G. Ridder (2003). Efficient estimation of average treatment effects using the estimated propensity score. *Econometrica 71*(4), 1161–1189.

Huling, J. D. and M. Yu (2021). Subgroup identification using the personalized package. *Journal of Statistical Software 98*(5), 1–60.

Iorio, F., T. A. Knijnenburg, D. J. Vis, G. R. Bignell, M. P. Menden, M. Schubert, et al. (2016). A landscape of pharmacogenomic interactions in cancer. *Cell 166*(3), 740–754.

Krivobokova, T., C. M. Crainiceanu, and G. Kauermann (2008). Fast adaptive penalized splines. *Journal of Computational and Graphical Statistics 17*(1), 1–20.

Liang, M., T. Ye, and H. Fu (2018). Estimating individualized optimal combination therapies through outcome weighted deep learning algorithms. *Statistics in Medicine 37*(27), 3869–3886.

Lin, Y. and H. H. Zhang (2006). Component selection and smoothing in multivariate nonparametric regression. *The Annals of Statistics 34*(5), 2272–2297.

Meier, L., S. Van de Geer, and P. Bühlmann (2009). High-dimensional additive modeling. *The Annals of Statistics 37*(6B), 3779–3821.

Mi, X., F. Zou, and R. Zhu (2019). Bagging and deep learning in optimal individualized treatment rules. *Biometrics 75*(2), 674–684.

Murdoch, W. J., C. Singh, K. Kumbier, R. Abbasi-Asl, and B. Yu (2019). Definitions, methods, and applications in interpretable machine learning. *Proceedings of the National Academy of Sciences 116*(44), 22071–22080.

Nie, X. and S. Wager (2021). Quasi-oracle estimation of heterogeneous treatment effects. *Biometrika 108*(2), 299–319. asaa076.

Pan, Y. and Y.-Q. Zhao (2021). Improved doubly robust estimation in learning optimal individualized treatment rules. *Journal of the American Statistical Association 116*(533), 283–294.

Park, H., E. Petkova, T. Tarpey, and R. T. Ogden (2022). A sparse additive model for treatment effect-modifier selection. *Biostatistics 23*(2), 412–429.

Petersen, A. and D. Witten (2019). Data-adaptive additive modeling. *Statistics in Medicine 38*(4), 583–600.

Petersen, A., D. Witten, and N. Simon (2016). Fused lasso additive model. *Journal of Computational and Graphical Statistics 25*(4), 1005–1025.

Qi, Z., D. Liu, H. Fu, and Y. Liu (2020). Multi-armed angle-based direct learning for estimating optimal individualized treatment rules with various outcomes. *Journal of the American Statistical Association 115*(530), 678–691.

Qian, M. and S. A. Murphy (2011). Performance guarantees for individualized treatment rules. *The Annals of Statistics 39*(2), 1180.

Ravikumar, P., J. Lafferty, H. Liu, and L. Wasserman (2009). Sparse additive models. *Journal of the Royal Statistical Society: Series B (Statistical Methodology) 71*(5), 1009–1030.

Rubin, D. B. (2005). Causal inference using potential outcomes: Design, modeling, decisions. *Journal of the American Statistical Association 100*(469), 322–331.

Rudin, C. (2019). Stop explaining black box machine learning models for high stakes decisions and use interpretable models instead. *Nature Machine Intelligence 1*(5), 206.

Shi, C., R. Song, and W. Lu (2021). Concordance and value information criteria for optimal treatment decision. *The Annals of Statistics 49*(1), 49 – 75.

Tay, J. K. and R. Tibshirani (2020). Reluctant generalized additive modeling. *International Statistical Review 88*(S1), S205–S224.

Tian, L., A. A. Alizadeh, A. J. Gentles, and R. Tibshirani (2014). A simple method for estimating interactions between a treatment and a large number of covariates. *Journal of the American Statistical Association 109*(508), 1517–1532.

Tibshirani, R. (1996). Regression shrinkage and selection via the lasso. *Journal of the Royal Statistical Society B 58*(1), 267–288.

Vayena, E., A. Blasimme, and I. G. Cohen (2018). Machine learning in medicine: Addressing ethical challenges. *PLoS medicine 15*(11), e1002689.

Wager, S. and S. Athey (2018). Estimation and inference of heterogeneous treatment effects using random forests. *Journal of the American Statistical Association 113*(523), 1228–1242.

Wahba, G. (2006). *Splines in nonparametric regression*, Volume 4. Wiley Online Library.

Wallace, M. P. and E. E. Moodie (2015). Doubly-robust dynamic treatment regimen estimation via weighted least squares. *Biometrics 71*(3), 636–644.

Yang, W., J. Soares, P. Greninger, E. J. Edelman, H. Lightfoot, S. Forbes, N. Bindal, D. Beare, J. A. Smith, I. R. Thompson, et al. (2012). Genomics of drug sensitivity in cancer (gdsc): a resource for therapeutic biomarker discovery in cancer cells. *Nucleic Acids Research 41*(D1), D955–D961.

Yu, G., J. Bien, and R. Tibshirani (2019). Reluctant interaction modeling. *arXiv preprint arXiv:1907.08414*.

Zhao, Y., D. Zeng, A. J. Rush, and M. R. Kosorok (2012). Estimating individualized treatment rules using outcome weighted learning. *Journal of the American Statistical Association 107*(499), 1106–1118.

Zhou, X., N. Mayer-Hamblett, U. Khan, and M. R. Kosorok (2017). Residual weighted learning for estimating individualized treatment rules. *Journal of the American Statistical Association 112*(517), 169–187.

Zhu, R., Y.-Q. Zhao, G. Chen, S. Ma, and H. Zhao (2017). Greedy outcome weighted tree learning of optimal personalized treatment rules. *Biometrics 73*(2), 391–400.

Supplementary material for "A reluctant additive
model framework for interpretable nonlinear
individualized treatment rules"
by Maronge, Huling, and Chen

November 2, 2023

# A  RAITR with Nuisance Parameters Estimated by Cross-fitting

Here we describe in full detail our proposed methodology when utilized with cross-fitting. This algorithm is the same as the main text, but with the cross-fitting procedure fully incorporated in full notation.

---

**Algorithm 2:** *Reluctant GAMs for Interpretable Nonlinear ITR with Nuisance Parameter Estimation*

---

**Input:** $(\mathbf{X}_i, Y_i, A_i)$, $i = 1, \ldots, n$: covariate $\mathbf{X}$ (with dimension $p$), outcome $Y$, treatment $A$, and penalty terms $\lambda_1, \lambda_2 \geq 0$.

**Process:**

1. Take a $K$-fold random partition $(I_k)_{k=1}^K$ Also, for each $k \in [K] = 1, ..., K$ , define the complement set $I_k^c := \{1, ..., N\} \backslash I_k$.

2. For each $k \in K$, using only data in using $I_k^c$, we construct the propensity score model $\hat{\pi}^{I_k^c}(\mathbf{X}, A)$ and main effects model $\hat{m}^{I_k^c}(\mathbf{X})$.

3. Fit the weighted learning with least square loss and linear rule:

$$\hat{\boldsymbol{\beta}}_1 = \operatorname{argmin}_{\boldsymbol{\beta}} \sum_{k=1}^K \sum_{i \in I^k} \hat{\pi}^{I_k^c}(\mathbf{X}_i, A_i)^{-1} \left\{ y_i - \hat{m}^{I_k^c}(\mathbf{X}_i) - \frac{A_i}{2} \boldsymbol{\beta}^T \mathbf{X}_i \right\}^2 + \lambda_1 \sum_{j=1}^p |\beta_j| \, ,$$

and compute residuals $r_i = y_i - m^{I_k^c}(\mathbf{X}_i) - (A_i/2)\hat{\boldsymbol{\beta}}_1^T \mathbf{X}_i, i = 1, \ldots n$.

4. For each covariate $j \in \{1, \ldots, p\}$, fit a weighted $(\hat{\pi}^{I_k^c}(\mathbf{X}_i, A_i)^{-1}$ as the weight) penalized smoothing spline (with data-dependent degrees of freedom) of $r_i$ on $g_j$.

5. Fit penalized regression with LASSO penalty of $A$ on $\mathbf{X}$ and $\mathbf{G}$:

$$\hat{\boldsymbol{\beta}} = \operatorname{argmin}_{\boldsymbol{\beta}} \sum_{k=1}^K \sum_{i \in I^k} \hat{\pi}^{I_k^c}(\mathbf{X}_i, A_i)^{-1} \left\{ y_i - \hat{m}^{I_k^c}(\mathbf{X}_i) - \frac{A_i}{2} \left( \boldsymbol{\beta}_{lin}^T \mathbf{X}_i + \boldsymbol{\beta}_{non}^T \mathbf{g}_i \right) \right\}^2$$
$$+ \lambda_2 \sum_{j=1}^p (|\boldsymbol{\beta}_{lin,j}| + \gamma_j |\boldsymbol{\beta}_{non,j}|) \, ,$$

with penalty $\gamma_j = \min(\sqrt{p}, 1 + s_j^{-1})$.

6. $\hat{f}(\mathbf{X}_i) = \hat{\boldsymbol{\beta}}_{lin} \mathbf{X}_i + \hat{\boldsymbol{\beta}}_{non} \mathbf{G}_i$ for $i = 1, \ldots, n$.

---

# B   Additional Simulation Details and Results

## B.1   Computational load

To study the computational load of each method, we consider two cases from our simulation settings: low sample size with moderate dimensional covariates and large sample size with high dimensional covariates. We generate data under the "small effect" highly nonlinear setting with $n = 250$, $p = 100$ and with $n = 1000$, $p = 1000$ and record the mean and standard deviation of the computation time across 20 replicates. We show results for RAITR, GRF, and samTEMsel. We split the computation time for RAITR into "Nuisance parameter fitting", which includes propensity score fitting and augmentation (i.e., Step 1 and Step 2 of Algorithm 2), and "RAITR CV" or "RAITR CIC", which includes constructing the nonlinear terms, final model fitting and ITR construction (i.e., Step 3 to Step 6 of Algorithm 2). We only show RAITR using cross-validation or CIC because the particular version of RAITR beyond the distinction of CIC or CV did not affect computation times significantly. The results are shown in Table B.1. Note that linear CV is approximately equal to RAITR prep and that a majority of computation time for our method is due to the augmentation procedure. All procedures were performed on a 2021 Macbook Pro with M1 Pro chip and 16 GB of RAM.

Our proposed method is more computationally intensive than the comparators; however, it is mainly due to the nuisance parameters being estimated using XGboost [1]. The computational time of our method could be substantially decreased if other models were used, such as Random Forests (quickly computable with fast implementations such as `ranger` [2]) or high dimensional parametric models including LASSO or Ridge.

Table B.1: Mean and standard deviation of computation time per simulation (in seconds) with RAITR and comparator methods across 20 replicates.

| Method | $n = 250$, $p = 100$ | $n = 1000$, $p = 1000$ |
|---|---|---|
| Nuisance parameter fitting with XGBoost | 66.2 (8.24) | 2257.9 (278.55) |
| RAITR CV | 2.9 (0.62) | 49.2 (4.77) |
| RAITR CIC | 1.0 (0.09) | 36.7 (1.98) |
| GRF | 0.4 (0.02) | 2.3 (0.12) |
| samTEMsel | 1.5 (0.06) | 23.8 (1.07) |

## B.2 Covariate balance

In this section, we give further details on the balance of covariates in the treated and control group for our data-generating model in Section 3. We only use one unique propensity score model in our simulations, as discussed in Section 3. To study the covariate imbalance under this propensity score model, we generate data under the "small effect" highly nonlinear setting with $n = 250$, $p = 100$, and with $n = 1000$, $p = 1000$ and record the absolute standardized mean difference and standard deviation across 500 replicates. In particular, $X_1$, $X_2$, and $X_3$ are confounders, and $X_4$ and $X_5$ are non-confounders. The results are shown in Table B.2.

Table B.2: Absolute standardized mean difference and proportion treated with associated standard deviations across 500 replicates corresponding to the propensity model described in Section 3.

|  | $n = 250,\ p = 100$ | $n = 1000,\ p = 1000$ |
|---|---|---|
| $X_1$ | 0.86 (0.14) | 0.86 (0.07) |
| $X_2$ | 0.85 (0.15) | 0.85 (0.07) |
| $X_3$ | 0.39 (0.13) | 0.38 (0.06) |
| $X_4$ | 0.10 (0.07) | 0.05 (0.04) |
| $X_5$ | 0.10 (0.07) | 0.05 (0.04) |
| Proportion treated | 0.50 (0.03) | 0.50 (0.02) |

## B.3 Strength of signal

In this section, we capture the strength of the interaction effects signal in the simulation studies shown in Section 3, where we present "small effects" and "large effects" (of the treatment and covariate interaction). These effects are defined by inspection of the ratio of the square root of the variance of the interaction effects divided by the square root of the sum of the variances of the main effects and random error. Specifically, we calculate

$$\sqrt{\frac{\mathrm{Var}(\Delta(\mathbf{X}))}{\mathrm{Var}(m(\mathbf{X})) + \mathrm{Var}(\epsilon)}}\ ,$$

where $\Delta(\mathbf{X})$ denotes the form of the interaction effects, $m(\mathbf{X})$ denotes the form of the main effects, and $\epsilon$ is the random error (with variance equal to 4 in our simulation study). This is presented for each data-generation model for both "small effect" and "large effect" in Table B.3. The small effects scenarios all have interaction effects that are much smaller in the magnitude of variation than the main effects, making the signal-to-noise ratio of the interaction effects quite small; these settings are thus especially challenging. The large effects scenarios have a larger signal-to-noise ratio than the small effects scenarios but still do not have extremely large signal-to-noise ratios.

Table B.3: The mean (sd) strength of the signal for nonlinear effects for "small effect" and "large effect" in the four data-generation methods described in Section 3 across 500 replicates. This quantity is captured by taking the ratio of the square root of the variance of the interaction effects divided by the square root of the sum of the variances of the main effects and the random error.

|  | "Small effect" | "Large effect" |
|---|---|---|
| Linear | 0.18 (0.002) | 1.20 (0.008) |
| Polynomial | 0.26 (0.003) | 1.16 (0.007) |
| Tree | 0.21 (0.002) | 1.14 (0.004) |
| Highly nonlinear | 0.07 (0.001) | 0.49 (0.004) |

## B.4   Additional simulation results

Here we show the results of value function from the simulation study described in Section 3 (with the value of "optimal" rule serving as the benchmark). The results are shown for each of the ITRs: 1. linear, 2. highly nonlinear, 3. tree, and 4. polynomial. Each ITR has a "large effect" and "small effect", as described in Section 3.

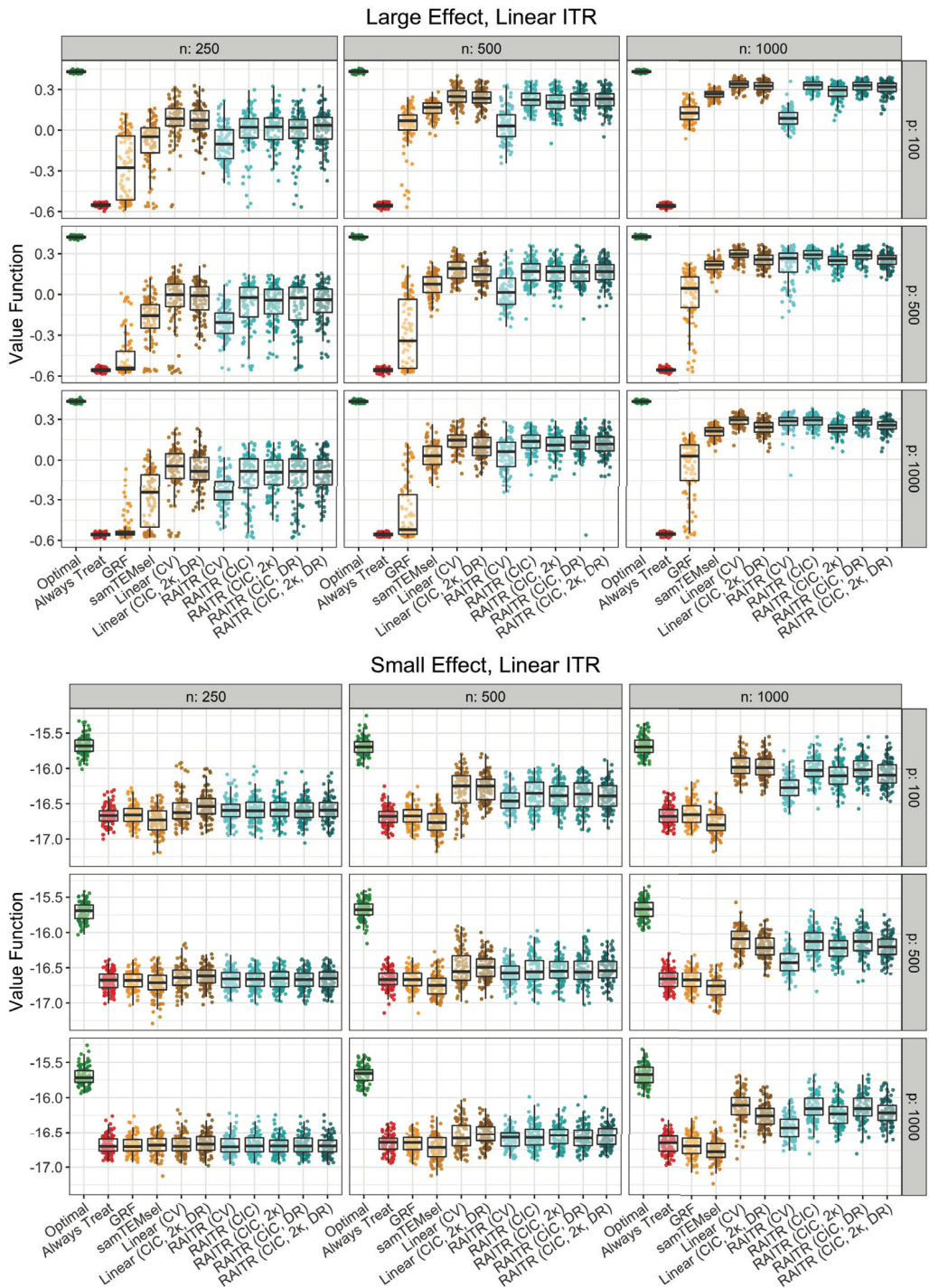

Figure B.1: Top: results for value function for large effect size, linear data-generating model for $n = 250,\ 500,\ 1000$ and $p = 100,\ 500,\ 1000$ for RAITR, linear ITRs, GRF, and samTEMsel. Bottom: as for top but small effect size.

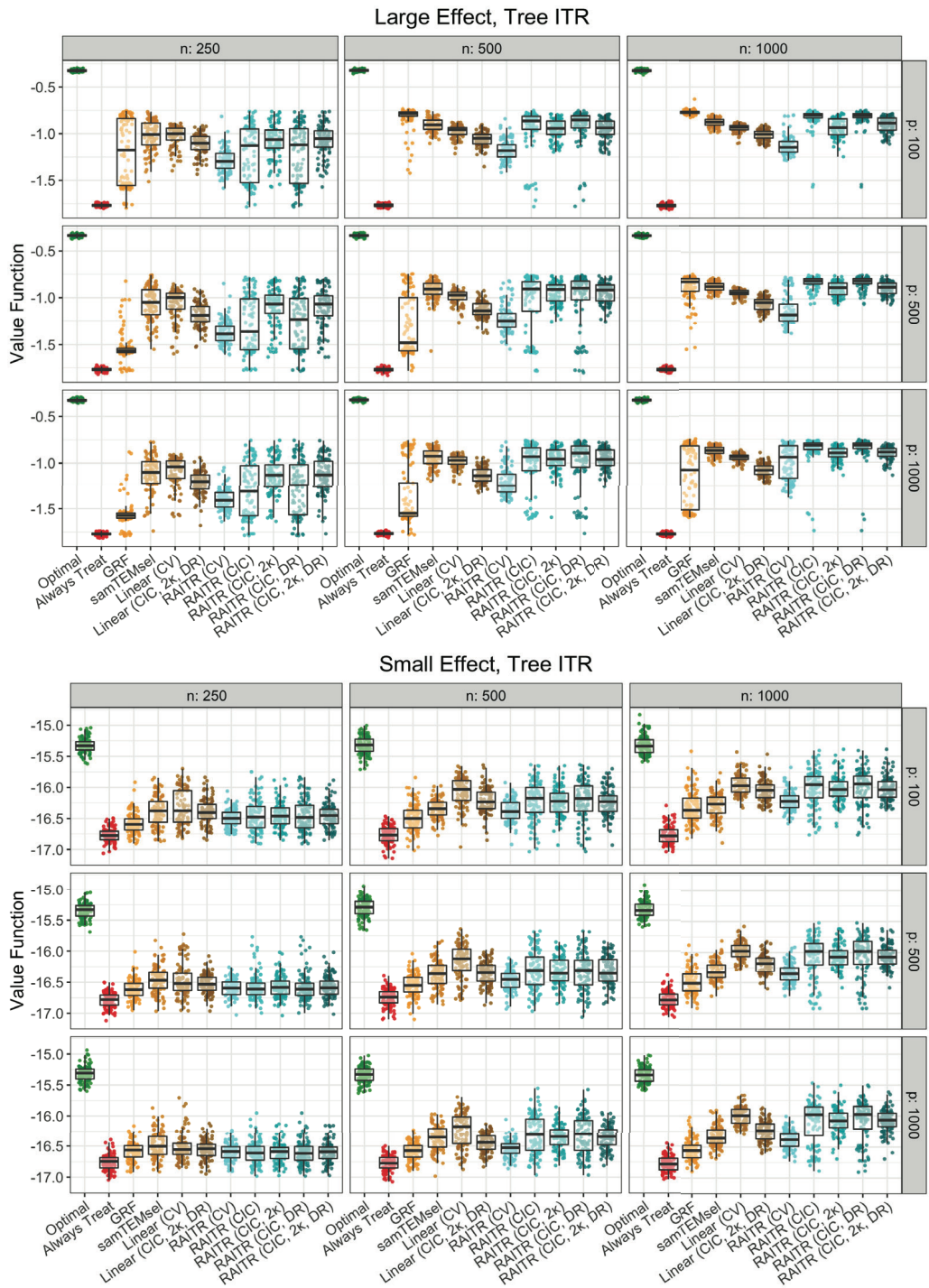

Figure B.2: Top: results for value function for large effect size, tree data-generating model for $n = 250,\ 500,\ 1000$ and $p = 100,\ 500,\ 1000$ for RAITR, linear ITRs, GRF, and samTEMsel. Bottom: as for top but small effect size.

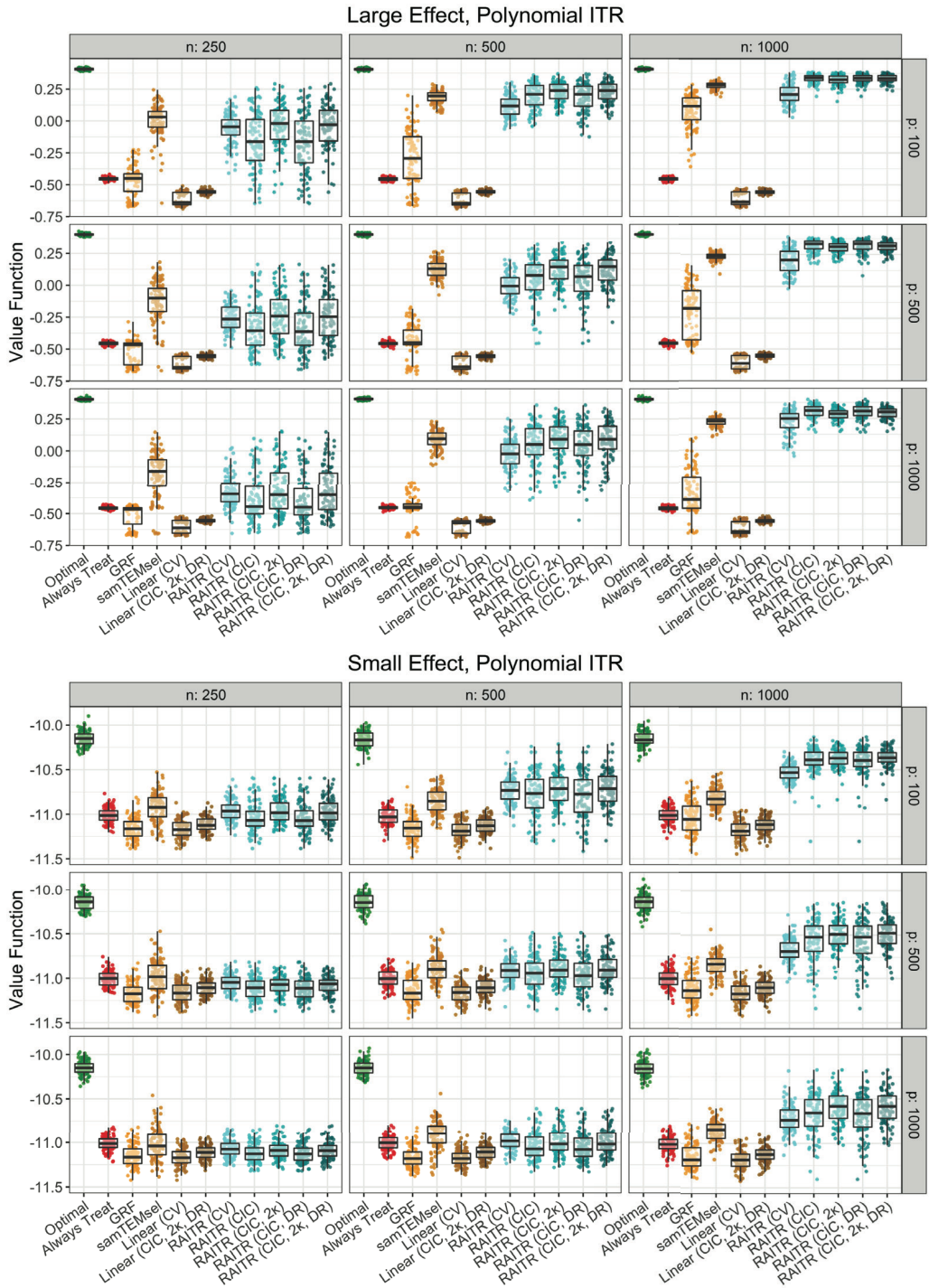

Figure B.3: Top: results for value function for large effect size, polynomial data-generating model for $n = 250$, $500$, $1000$ and $p = 100$, $500$, $1000$ for RAITR, linear ITRs, GRF, and samTEMsel. Bottom: as for top but small effect size.

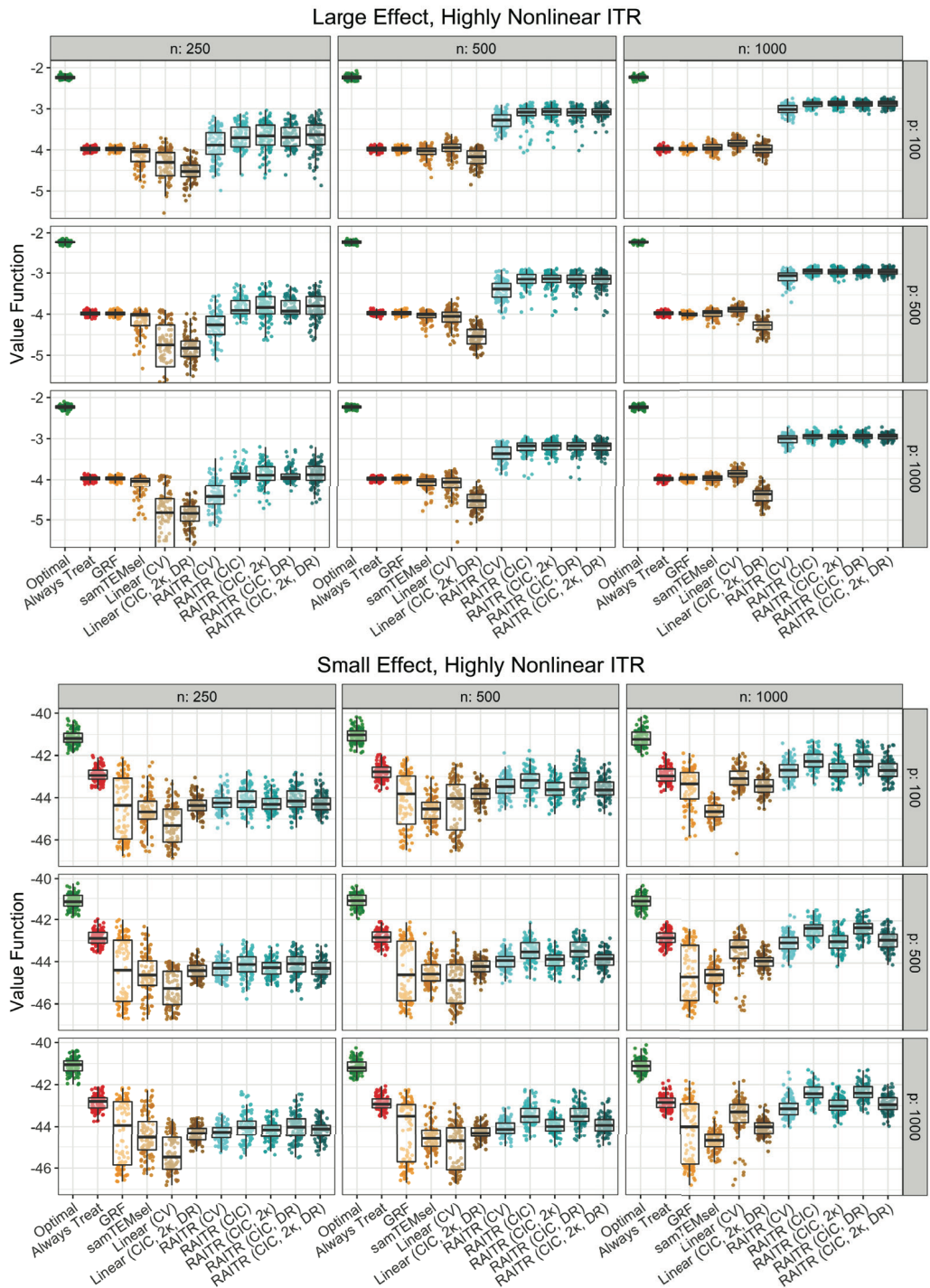

Figure B.4: Top: results for value function for large effect size, highly nonlinear data-generating model for $n = 250, 500, 1000$ and $p = 100, 500, 1000$ for RAITR, linear ITRs, GRF, and samTEMsel. Bottom: as for top but small effect size.

# C Additional results using GDSC data

In Table C.1 we show the drug names for each pair number, as well as the sample size. In Tables C.2 and C.3, we show the agreement and value results from 26 pairs of drugs from the GDSC study. Similar to the results we have shown in the main text, RAITR, particularly RAITR (CIC, $2\kappa$, DR), performs well across all drug pairs.

Table C.1: Drug names for drug pair number and sample size ($n$) for each drug pair in GDSC analysis. All analyses use 1486 genes.

| Drug pair number | Drugs | $n$ |
|---|---|---|
| 1 | SN-38 and Vinblastine | 927 |
| 2 | SN-38 and AZD4877 | 915 |
| 3 | Omipalisib and Temsirolimus | 856 |
| 4 | Temsirolimus and Doxorubicin | 909 |
| 5 | Doxorubicin and PD0325901 | 912 |
| 6 | Doxorubicin and rTRAIL | 902 |
| 7 | AZD7762 and IGFR_3801 | 913 |
| 8 | AZD2014 and IOX2 | 902 |
| 9 | IOX2 and AZD8055 | 893 |
| 10 | IOX2 and IGFR_3801 | 898 |
| 11 | AZD8055 and IGFR_3801 | 894 |
| 12 | AZD8055 and Cytarabine | 902 |
| 13 | IGFR_3801 and CD532 | 868 |
| 14 | PARP_0108 and CD532 | 869 |
| 15 | CD532 and WZ3105 | 881 |
| 16 | Cytarabine and IMD-0354 | 866 |
| 17 | IMD-0354 and Gefitinib | 869 |
| 18 | Gefitinib and Vorinostat | 922 |
| 19 | JQ1 and Navitoclax | 921 |
| 20 | Refametinib and PARP_9495 | 907 |
| 21 | Refametinib and AZD6738 | 914 |
| 22 | Piperlongumine and AZD6738 | 899 |
| 23 | AZD6738 and AZD5582 | 902 |
| 24 | AZD5582 and Bosutinib | 894 |
| 25 | Bosutinib and AZD8931 | 902 |
| 26 | Bleomycin (10 $\mu$M) and AZD1332 | 906 |

Table C.2: Extended agreement results from GDSC analysis for all drugs. Headers L. and R. reduced to increase table size. These denote linear and RAITR, respectively. The entries give mean agreement over 100 replicates with standard deviation in parentheses.

| Pair | Always Treat | GRF | samTEMsel | Linear (CV) | L. (CIC, $2\kappa$, DR) | RAITR (CV) | RAITR (CIC) | R. (CIC, $2\kappa$) | R. (CIC, DR) | R. (CIC, $2\kappa$, DR) |
|---|---|---|---|---|---|---|---|---|---|---|
| 1 | 0.52 (0.022) | 0.53 (0.030) | 0.56 (0.043) | 0.63 (0.048) | 0.63 (0.034) | 0.61 (0.037) | 0.60 (0.052) | 0.63 (0.038) | 0.60 (0.054) | 0.63 (0.038) |
| 2 | 0.65 (0.020) | 0.65 (0.020) | 0.66 (0.022) | 0.67 (0.062) | 0.67 (0.034) | 0.67 (0.030) | 0.65 (0.033) | 0.68 (0.029) | 0.65 (0.041) | 0.68 (0.044) |
| 3 | 0.64 (0.023) | 0.64 (0.022) | 0.66 (0.022) | 0.65 (0.058) | 0.65 (0.030) | 0.66 (0.026) | 0.65 (0.029) | 0.66 (0.030) | 0.65 (0.026) | 0.66 (0.025) |
| 4 | 0.57 (0.023) | 0.57 (0.023) | 0.57 (0.022) | 0.57 (0.039) | 0.58 (0.027) | 0.57 (0.025) | 0.57 (0.024) | 0.58 (0.027) | 0.57 (0.026) | 0.58 (0.027) |
| 5 | 0.66 (0.024) | 0.67 (0.026) | 0.69 (0.024) | 0.69 (0.053) | 0.69 (0.046) | 0.70 (0.028) | 0.67 (0.027) | 0.70 (0.025) | 0.67 (0.027) | 0.70 (0.025) |
| 6 | 0.72 (0.023) | 0.72 (0.023) | 0.72 (0.021) | 0.70 (0.037) | 0.70 (0.030) | 0.71 (0.026) | 0.72 (0.024) | 0.71 (0.027) | 0.72 (0.024) | 0.71 (0.030) |
| 7 | 0.60 (0.028) | 0.60 (0.028) | 0.60 (0.023) | 0.58 (0.055) | 0.59 (0.026) | 0.60 (0.026) | 0.60 (0.036) | 0.60 (0.030) | 0.60 (0.029) | 0.60 (0.029) |
| 8 | 0.54 (0.023) | 0.53 (0.040) | 0.53 (0.041) | 0.53 (0.041) | 0.55 (0.028) | 0.54 (0.037) | 0.54 (0.036) | 0.55 (0.031) | 0.54 (0.037) | 0.55 (0.030) |
| 9 | 0.50 (0.025) | 0.56 (0.032) | 0.55 (0.025) | 0.55 (0.036) | 0.56 (0.030) | 0.55 (0.032) | 0.53 (0.036) | 0.56 (0.031) | 0.53 (0.035) | 0.56 (0.028) |
| 10 | 0.49 (0.022) | 0.50 (0.036) | 0.54 (0.039) | 0.59 (0.035) | 0.58 (0.028) | 0.58 (0.036) | 0.52 (0.040) | 0.56 (0.038) | 0.53 (0.042) | 0.57 (0.039) |
| 11 | 0.48 (0.025) | 0.59 (0.033) | 0.61 (0.030) | 0.60 (0.038) | 0.60 (0.032) | 0.59 (0.034) | 0.56 (0.042) | 0.59 (0.037) | 0.56 (0.043) | 0.60 (0.036) |
| 12 | 0.52 (0.023) | 0.55 (0.039) | 0.54 (0.030) | 0.57 (0.037) | 0.58 (0.030) | 0.57 (0.035) | 0.54 (0.037) | 0.57 (0.035) | 0.54 (0.037) | 0.58 (0.033) |
| 13 | 0.60 (0.025) | 0.65 (0.029) | 0.66 (0.024) | 0.65 (0.042) | 0.66 (0.026) | 0.66 (0.035) | 0.65 (0.044) | 0.67 (0.029) | 0.65 (0.038) | 0.67 (0.029) |
| 14 | 0.52 (0.021) | 0.65 (0.031) | 0.68 (0.026) | 0.70 (0.046) | 0.70 (0.040) | 0.68 (0.037) | 0.69 (0.041) | 0.70 (0.036) | 0.69 (0.044) | 0.71 (0.035) |
| 15 | 0.46 (0.023) | 0.58 (0.045) | 0.55 (0.050) | 0.61 (0.033) | 0.62 (0.031) | 0.60 (0.035) | 0.56 (0.046) | 0.60 (0.036) | 0.57 (0.043) | 0.61 (0.035) |
| 16 | 0.54 (0.025) | 0.61 (0.034) | 0.58 (0.035) | 0.60 (0.036) | 0.61 (0.029) | 0.60 (0.034) | 0.59 (0.033) | 0.61 (0.026) | 0.59 (0.032) | 0.61 (0.024) |
| 17 | 0.67 (0.023) | 0.69 (0.029) | 0.72 (0.027) | 0.71 (0.032) | 0.71 (0.035) | 0.71 (0.033) | 0.68 (0.030) | 0.71 (0.029) | 0.69 (0.032) | 0.71 (0.028) |
| 18 | 0.53 (0.025) | 0.63 (0.034) | 0.64 (0.023) | 0.65 (0.028) | 0.65 (0.030) | 0.64 (0.027) | 0.58 (0.052) | 0.64 (0.033) | 0.59 (0.052) | 0.64 (0.031) |
| 19 | 0.59 (0.025) | 0.66 (0.025) | 0.65 (0.025) | 0.62 (0.042) | 0.62 (0.027) | 0.63 (0.033) | 0.63 (0.042) | 0.64 (0.032) | 0.63 (0.039) | 0.64 (0.031) |
| 20 | 0.46 (0.023) | 0.65 (0.050) | 0.67 (0.024) | 0.68 (0.027) | 0.68 (0.026) | 0.66 (0.036) | 0.65 (0.041) | 0.68 (0.028) | 0.65 (0.040) | 0.68 (0.028) |
| 21 | 0.50 (0.024) | 0.68 (0.026) | 0.69 (0.027) | 0.70 (0.026) | 0.70 (0.025) | 0.69 (0.029) | 0.68 (0.037) | 0.70 (0.026) | 0.68 (0.026) | 0.70 (0.026) |
| 22 | 0.52 (0.023) | 0.59 (0.038) | 0.54 (0.039) | 0.58 (0.054) | 0.60 (0.033) | 0.59 (0.042) | 0.56 (0.044) | 0.60 (0.038) | 0.57 (0.043) | 0.60 (0.035) |
| 23 | 0.58 (0.020) | 0.57 (0.060) | 0.56 (0.047) | 0.59 (0.057) | 0.60 (0.031) | 0.60 (0.032) | 0.56 (0.052) | 0.58 (0.040) | 0.56 (0.052) | 0.58 (0.041) |
| 24 | 0.47 (0.024) | 0.52 (0.048) | 0.56 (0.042) | 0.55 (0.031) | 0.56 (0.027) | 0.56 (0.028) | 0.53 (0.033) | 0.55 (0.030) | 0.53 (0.033) | 0.55 (0.030) |
| 25 | 0.48 (0.023) | 0.65 (0.027) | 0.63 (0.023) | 0.64 (0.029) | 0.64 (0.028) | 0.63 (0.032) | 0.61 (0.047) | 0.64 (0.028) | 0.61 (0.046) | 0.64 (0.026) |
| 26 | 0.50 (0.023) | 0.59 (0.033) | 0.58 (0.032) | 0.62 (0.035) | 0.63 (0.030) | 0.61 (0.033) | 0.59 (0.036) | 0.62 (0.031) | 0.60 (0.034) | 0.62 (0.031) |

Table C.3: Value results from GDSC analysis for all drugs. Headers L. and R. reduced to increase table size. These denote linear and RAITR, respectively. The entries give the mean value function over 100 replicates with standard deviation in parentheses.

| Pair | Optimal | Always Treat | GRF | samTEMsel | Linear (CV) | L. (CIC, 2s, DR) | RAITR (CV) | RAITR (CIC) | R. (CIC, 2s) | R. (CIC, DR) | R. (CIC, 2s, DR) |
|---|---|---|---|---|---|---|---|---|---|---|---|
| 1 | 4.82 (0.069) | 4.27 (0.073) | 4.28 (0.085) | 4.32 (0.088) | 4.39 (0.110) | 4.40 (0.083) | 4.38 (0.081) | 4.37 (0.094) | 4.40 (0.090) | 4.37 (0.096) | 4.40 (0.090) |
| 2 | 4.63 (0.077) | 4.27 (0.087) | 4.27 (0.087) | 4.25 (0.078) | 4.25 (0.150) | 4.27 (0.100) | 4.28 (0.090) | 4.27 (0.097) | 4.29 (0.090) | 4.26 (0.120) | 4.29 (0.120) |
| 3 | 3.42 (0.078) | 2.87 (0.089) | 2.87 (0.089) | 2.90 (0.095) | 2.87 (0.160) | 2.88 (0.098) | 2.89 (0.087) | 2.87 (0.100) | 2.90 (0.100) | 2.88 (0.092) | 2.91 (0.091) |
| 4 | 2.56 (0.059) | 2.09 (0.070) | 2.09 (0.070) | 2.09 (0.070) | 2.05 (0.100) | 2.06 (0.074) | 2.07 (0.074) | 2.09 (0.069) | 2.08 (0.079) | 2.09 (0.076) | 2.08 (0.080) |
| 5 | 2.24 (0.064) | 1.68 (0.065) | 1.68 (0.067) | 1.76 (0.075) | 1.76 (0.140) | 1.76 (0.120) | 1.78 (0.084) | 1.71 (0.073) | 1.79 (0.074) | 1.71 (0.072) | 1.79 (0.074) |
| 6 | 2.04 (0.058) | 1.67 (0.060) | 1.67 (0.060) | 1.69 (0.070) | 1.63 (0.090) | 1.64 (0.072) | 1.66 (0.068) | 1.67 (0.061) | 1.66 (0.070) | 1.66 (0.063) | 1.65 (0.073) |
| 7 | 0.81 (0.063) | 0.44 (0.071) | 0.44 (0.071) | 0.41 (0.066) | 0.37 (0.140) | 0.39 (0.078) | 0.42 (0.073) | 0.43 (0.096) | 0.43 (0.073) | 0.43 (0.075) | 0.43 (0.075) |
| 8 | 0.61 (0.041) | 0.08 (0.055) | 0.12 (0.061) | 0.10 (0.064) | 0.13 (0.055) | 0.15 (0.052) | 0.14 (0.059) | 0.13 (0.054) | 0.15 (0.052) | 0.13 (0.056) | 0.15 (0.053) |
| 9 | 0.47 (0.038) | 0.06 (0.035) | 0.13 (0.048) | 0.11 (0.042) | 0.09 (0.068) | 0.11 (0.051) | 0.12 (0.052) | 0.09 (0.057) | 0.12 (0.050) | 0.09 (0.052) | 0.12 (0.048) |
| 10 | 0.64 (0.048) | 0.07 (0.033) | 0.08 (0.040) | 0.09 (0.079) | 0.14 (0.097) | 0.15 (0.065) | 0.15 (0.066) | 0.09 (0.057) | 0.13 (0.054) | 0.10 (0.058) | 0.14 (0.062) |
| 11 | 0.59 (0.049) | -0.06 (0.042) | 0.12 (0.071) | 0.16 (0.058) | 0.13 (0.097) | 0.14 (0.074) | 0.12 (0.080) | 0.06 (0.081) | 0.13 (0.076) | 0.06 (0.083) | 0.13 (0.074) |
| 12 | 0.48 (0.057) | -0.05 (0.051) | -0.01 (0.065) | -0.02 (0.064) | 0.01 (0.088) | 0.03 (0.067) | 0.01 (0.073) | -0.02 (0.075) | 0.02 (0.072) | -0.02 (0.073) | 0.03 (0.071) |
| 13 | 0.56 (0.046) | -0.08 (0.078) | 0.14 (0.098) | 0.14 (0.062) | 0.16 (0.080) | 0.17 (0.062) | 0.15 (0.071) | 0.11 (0.110) | 0.17 (0.065) | 0.11 (0.096) | 0.17 (0.067) |
| 14 | 0.65 (0.062) | -0.14 (0.085) | 0.16 (0.086) | 0.25 (0.075) | 0.26 (0.110) | 0.27 (0.097) | 0.23 (0.092) | 0.25 (0.110) | 0.27 (0.094) | 0.25 (0.100) | 0.28 (0.080) |
| 15 | 0.44 (0.055) | -0.21 (0.047) | -0.05 (0.096) | -0.11 (0.100) | -0.03 (0.083) | -0.02 (0.071) | -0.04 (0.076) | -0.10 (0.100) | -0.04 (0.079) | -0.09 (0.094) | -0.03 (0.078) |
| 16 | 0.46 (0.055) | -0.21 (0.072) | -0.06 (0.083) | -0.13 (0.090) | -0.06 (0.079) | -0.05 (0.070) | -0.07 (0.089) | -0.09 (0.078) | -0.04 (0.075) | -0.09 (0.073) | -0.04 (0.072) |
| 17 | 0.06 (0.051) | -0.40 (0.065) | -0.34 (0.086) | -0.27 (0.061) | -0.26 (0.074) | -0.26 (0.074) | -0.28 (0.077) | -0.35 (0.086) | -0.27 (0.076) | -0.34 (0.091) | -0.27 (0.074) |
| 18 | -0.48 (0.046) | -0.90 (0.052) | -0.76 (0.061) | -0.75 (0.054) | -0.75 (0.059) | -0.76 (0.062) | -0.76 (0.056) | -0.83 (0.075) | -0.76 (0.064) | -0.83 (0.081) | -0.76 (0.063) |
| 19 | -0.44 (0.070) | -1.10 (0.055) | -0.89 (0.075) | -0.90 (0.081) | -0.94 (0.100) | -0.93 (0.085) | -0.92 (0.086) | -0.96 (0.110) | -0.90 (0.085) | -0.96 (0.110) | -0.90 (0.083) |
| 20 | -0.95 (0.070) | -1.62 (0.082) | -1.31 (0.100) | -1.28 (0.072) | -1.28 (0.081) | -1.28 (0.081) | -1.31 (0.088) | -1.34 (0.097) | -1.29 (0.084) | -1.34 (0.095) | -1.29 (0.085) |
| 21 | -0.91 (0.061) | -1.62 (0.075) | -1.23 (0.065) | -1.22 (0.068) | -1.21 (0.064) | -1.21 (0.070) | -1.22 (0.070) | -1.24 (0.081) | -1.21 (0.069) | -1.24 (0.078) | -1.21 (0.069) |
| 22 | -1.25 (0.035) | -1.67 (0.037) | -1.58 (0.052) | -1.65 (0.061) | -1.58 (0.067) | -1.56 (0.049) | -1.58 (0.057) | -1.61 (0.057) | -1.57 (0.054) | -1.61 (0.057) | -1.56 (0.053) |
| 23 | -0.91 (0.082) | -1.70 (0.058) | -1.58 (0.120) | -1.65 (0.094) | -1.48 (0.120) | -1.47 (0.10) | -1.47 (0.096) | -1.59 (0.11) | -1.51 (0.110) | -1.58 (0.110) | -1.51 (0.100) |
| 24 | -0.90 (0.085) | -1.73 (0.100) | -1.66 (0.120) | -1.62 (0.120) | -1.61 (0.120) | -1.55 (0.091) | -1.56 (0.090) | -1.64 (0.10) | -1.59 (0.100) | -1.64 (0.100) | -1.59 (0.100) |
| 25 | -1.15 (0.061) | -1.79 (0.059) | -1.51 (0.075) | -1.53 (0.068) | -1.52 (0.076) | -1.51 (0.070) | -1.53 (0.075) | -1.57 (0.091) | -1.52 (0.070) | -1.56 (0.090) | -1.51 (0.066) |
| 26 | -1.29 (0.071) | -1.92 (0.086) | -1.75 (0.091) | -1.77 (0.080) | -1.72 (0.090) | -1.71 (0.077) | -1.74 (0.086) | -1.76 (0.080) | -1.72 (0.076) | -1.75 (0.082) | -1.72 (0.076) |

# References

[1] Tianqi Chen and Carlos Guestrin. Xgboost: A scalable tree boosting system. In *Proceedings of the 22nd acm sigkdd international conference on knowledge discovery and data mining*, pages 785–794, 2016.

[2] Marvin N Wright and Andreas Ziegler. ranger: A fast implementation of random forests for high dimensional data in c++ and r. *arXiv preprint arXiv:1508.04409*, 2015.